\documentclass[preprint]{emulateapj}
\usepackage{graphicx}
\usepackage{amsmath}

\begin{document}

\title{Signatures of $\Lambda$CDM substructure in tidal debris}
\author{Jennifer M. Siegal-Gaskins\altaffilmark{1,2} and Monica Valluri\altaffilmark{1,3,4}}
\altaffiltext{1}{Kavli Institute for Cosmological Physics}
\altaffiltext{2}{Department of Physics, University of Chicago, 5640 S. Ellis
Avenue, Chicago, IL 60637 \\ 
{\tt jsg@kicp.uchicago.edu}
}
\altaffiltext{3}{Department of Astronomy and Astrophysics, University of Chicago, 5640 S. Ellis
Avenue, Chicago, IL 60637
}
\altaffiltext{4}{Department of Astronomy, University of Michigan, 500 Church Street, Ann Arbor, MI 48109\\
{\tt mvalluri@umich.edu}}

\begin{abstract}
In the past decade, surveys of the stellar component of the Galaxy have revealed a number of streams from tidally disrupted dwarf galaxies and globular clusters.  Simulations of hierarchical structure formation in $\Lambda$CDM cosmologies predict that the dark matter halo of a galaxy like the Milky Way contains hundreds of subhalos with masses of $\sim10^{8}$ M$_{\odot}$ and greater, and it has been suggested that the existence of coherent tidal streams is incompatible with the expected abundance of substructure.  We investigate the effects of dark matter substructure on tidal streams by simulating the disruption of a self-gravitating satellite on a wide range of orbits in different host models both with and without substructure.  We find that the halo shape and the specific orbital path more strongly determine the overall degree of disruption of the satellite than does the presence or absence of substructure, i.e., the changes in the large-scale properties of the tidal debris due to substructure are small compared to variations in the debris from different orbits in a smooth potential.  Substructure typically leads to an increase in the degree of clumpiness of the tidal debris in sky projection, and in some cases a more compact distribution in line-of-sight velocity.  Substructure also leads to differences in the location of sections of debris compared to the results of the smooth halo model, which may have important implications for the interpretation of observed tidal streams.  A unique signature of the presence of substructure in the halo which may be detectable by upcoming surveys is identified.  We conclude, however, that predicted levels of substructure are consistent with a detection of a coherent tidal stream from a dwarf galaxy.  \end{abstract}


\section{Introduction}
\label{sec:intro}

In recent years, surveys of the stellar component of the Milky Way 
 have revealed rich structure and a number of new tidal streams \citep{belokurov_etal_06field,grillmair_06,grillmair_dionatos_06cold,belokurov_etal_07orphan}.  
In addition to the well-studied Sagittarius (Sgr) stream \citep{ibata_irwin_lewis_stolte_01, majewski_etal_03, newberg_etal_03, martinez-delgado_etal_04} and the Monoceros stream \citep{newberg_etal_02, yanny_etal_03, penarrubia_etal_05}, streams from the tidal disruption of globular clusters such as Palomar 5 \citep{odenkirchen_etal_01, rockosi_etal_02, odenkirchen_etal_03, grillmair_dionatos_06pal} and NGC 5466 \citep{grillmair_johnson_06ngc5466, belokurov_etal_06ngc} have also been identified.  
Tidal streams probe the gravitational potential on large scales, and consequently may prove to be a valuable tool for constraining the mass distribution of the Galaxy.

Simulations of structure formation in $\Lambda$CDM cosmologies  predict as many as several hundred dense clumps with masses $\gtrsim 10^8$ M$_{\odot}$ in the halo, roughly an order of magnitude more than the observed abundance of dwarf galaxies surrounding the Milky Way \citep{klypin_etal_99, moore_etal_99}.  
One class of solutions to this `missing satellites' problem involves modifications of the properties of the dark matter particle, e.g., warm dark matter \citep{hogan_dalcanton_00} or self-interacting dark matter \citep{spergel_steinhardt_00}, in order to reduce the amount of substructure.  Suppression in the small-scale end of the primordial power spectrum has also been invoked to reduce the predicted abundance of subhalos \citep{kamionkowski_liddle_00, zentner_bullock_03}.  Alternatively, mechanisms by which only a fraction of the dark matter subhalos have visible stellar components at the present epoch have been proposed to explain this discrepancy \citep{bullock_kravtsov_weinberg_01,kravtsov_gnedin_klypin_04}.  In these scenarios, a large number of dark substructures are predicted to populate the halo.  

In the past few years the number of known satellites of the Milky Way has increased dramatically with the discovery of 11 new, faint objects which are likely to be dwarf galaxies or globular clusters \citep{willman_etal_05glob, willman_etal_05ursa, zucker_etal_06ursa, zucker_etal_06canes,belokurov_etal_06bootes, belokurov_etal_07cats, irwin_etal_07}, hinting that a significant number of very faint or dark satellites could populate the halo.
If such a population does exist, these structures may be detectable in several ways.  Gamma rays from the annihilation of dark matter particles in substructures may reveal their presence \citep[e.g.,][]{bergstrom_etal_99, calcaneo-roldan_moore_00, tasitsiomi_olinto_02, stoehr_etal_03, koushiappas_etal_04, diemand_kuhlen_madau_07}.  
Dark substructures may also be detectable 
via flux anomalies in multi-image gravitational lens systems \citep{mao_schneider_98, metcalf_madau_01, dalal_kochanek_02, chiba_02} or via spectroscopic lensing \citep{moustakas_metcalf_03}.  However, recent work favors microlensing by stars in the lens galaxy over dark matter substructure as an explanation for flux anomalies  \citep{pooley_blackburne_rappaport_etal_07}. 

In this study we take a theoretical approach to determine whether substructure as predicted by $\Lambda$CDM models may be detectable by its influence on the formation and properties of tidal streams from dwarf galaxies.  We select a variety of orbits in several host galaxy models and simulate the tidal disruption of a progenitor satellite on these orbits both with and without dark matter substructure to evaluate the influences of the orbital path, host potential, and presence of substructure on the tidal debris.  
In \S\ref{sec:orbits} we discuss the phase space properties of orbits in the context of tidal disruption, describe our host galaxy models, and outline our method for selecting a diverse range of orbits.  We give the details of our numerical simulations in \S\ref{sec:sims}, and present our results in \S\ref{sec:results}.  We discuss our study in relation to previous work on this subject in \S\ref{sec:disc}, and finally summarize our conclusions in \S\ref{sec:conc}.

\section{Orbits}
\label{sec:orbits}

\subsection{Phase space properties of orbits}
\label{sec:phase}
 
In this study we use some basic principles governing the motion of orbits in phase space to select a range of orbits which may lead to streams with diverse properties.  The structure of phase space in Hamiltonian dynamics is well studied, and we refer the reader to some of the many excellent texts on this subject \citep[e.g.,][]{binney_tremaine_87, lichtenburg_liberman_92} for a more thorough review than is presented here.

In a spherically symmetric potential, orbits conserve four integrals of motion (the energy, as well as each of the three components of angular momentum), so orbits in such potentials are 
generically rosettes confined to a single plane in configuration space.    
In a non-spherical potential most orbits conserve three integrals of motion ({\it regular} orbits) or fewer ({\it chaotic} orbits), and thus densely fill 3-dimensional volumes in configuration space.
 
A {\it resonant} orbit is a regular orbit whose fundamental oscillation frequencies in configuration space are related via a linear equation such as $l\omega_1+m\omega_2+n\omega_3 = 0$, where $l,m,n$ are integers. Such a relation between the fundamental frequencies of motion implies that only two of the three frequencies are independent, and is similar to requiring that a fourth integral of motion be conserved.  Consequently, a resonant orbit satisfying one or two resonance equations as above will be confined to a sheet or a closed loop, respectively, in configuration space \citep{valluri_merritt_98,merritt_valluri_99}. 

In most potentials that are near-integrable, the vast majority of orbits are regular orbits. Only a tiny fraction of phase space is expected to be occupied by chaotic orbits and resonant orbits \citep{lichtenburg_liberman_92}.  More numerous than the resonant or chaotic orbits, however, are the regular orbits which are `resonantly trapped'.  These orbits are confined to an island in phase space around a resonance and typically occupy only thin volumes in configuration space, and so tidal debris from satellites on these orbits would be similarly confined.  Furthermore, if a satellite's orbit is associated with a stable resonance, the tidal debris along the orbit may be less susceptible to perturbations by substructure than streams formed along non-resonant orbits.  Theoretical work has shown that the density of streams on resonant orbits decreases more slowly than on non-resonant orbits \citep{vogelsberger_white_helmi_etal_08}.

Despite the differences in the phase space properties of chaotic and regular orbits, it is generally not possible to distinguish between these classes without full orbit information for $\gtrsim 25$ dynamical times \citep[e.g.,][]{papaphilippou_laskar_96,valluri_merritt_98}, far longer than the expected age of detected tidal streams.
For this reason, we do not distinguish between chaotic and regular orbits, and instead consider only whether a given orbit is resonantly trapped.

\subsection{Orbit selection}
\label{sec:select}
We construct libraries of orbits using a method similar to that of \citet{cretton_etal_99}, and select from these libraries orbits on which to initialize the progenitor satellite in our numerical simulations.  Each library consists of orbits with a fixed energy $E$ and $z$-component of angular momentum $L_{z}$.  Fixing these two quantities in a given potential determines a set of coordinates which are turning points for orbits (where $v_{R}=v_{z}=0$), termed the zero velocity curve (ZVC).  A library is generated by initializing orbits from various positions on the ZVC, and integrating these orbits in the host potential.

Our host potential consists of a smooth NFW dark matter halo~\citep{navarro_etal_97}, a Miyamoto-Nagai disk~\citep{miyamoto_nagai_75}, and a Hernquist bulge~\citep{hernquist_90}.  The masses of these components and halo concentration are adopted from the favored model of \citet{klypin_zhao_somerville_02} which gives the halo mass $M_{\rm halo}=10^{12}$ M$_{\odot}$ and concentration $c=12$, the disk mass $M_{d}=4\times10^{10}$ M$_{\odot}$, and the bulge mass \citep[$m_{1}+m_{2}$ in][]{klypin_zhao_somerville_02} $M_{b}=8\times10^{9}$ M$_{\odot}$.
The scale lengths of the Miyamoto-Nagai disk, $a=6.5$ kpc and $b=0.26$ kpc, and the Hernquist bulge scale parameter, $a=$ 0.7 kpc, are taken from \citet{johnston_etal_96}. 
\citet{klypin_zhao_somerville_02} use exponential density profiles for the baryonic components of the Galaxy, but since these models are considerably more computationally expensive to implement in our simulations, we instead use the Miyamoto-Nagai disk and Hernquist bulge to describe the baryon distribution.  These parameters produce a rotation curve which is qualitatively similar, but not identical, to that of \citet{klypin_zhao_somerville_02}.  

Tidal debris can be dispersed both by precession of the orbit of the progenitor due to a non-spherical potential and by heating from interactions with substructure  \citep{ibata_lewis_irwin_quinn_02,mayer_etal_02}, so it is necessary to evaluate the relative importance of these two influences.  $\Lambda$CDM simulations of structure formation find that galaxy-sized dark matter halos are generally triaxial, although the shape of the halo may be radially dependent \citep[e.g.,][]{frenk_etal_88, kazantzidis_etal_04shapes, hayashi_etal_07}.  We consider the simpler case of an axisymmetric dark matter halo with the axis ratio independent of radius.
The lengths of the principal axes are defined to be $(x:y:z) = (a:b:c)$ with the $x$ and $y$ axes equal and the $z$-coordinate aligned with the axis of rotation of the Galaxy.  The axis ratio $q=c/a=c/b$ is varied  to generate orbit libraries for halos with $q=1$ (spherical), $q=1.3$ (prolate), and $q=0.6$ (oblate).  The potentials for the oblate and prolate versions of the NFW  halo are obtained by defining a variable $\tilde{r}$ which replaces $r$ in the spherically-symmetric NFW potential $\Phi_{\rm NFW}(r)$, where $\tilde{r}^{2}={x^{2}+y^{2}+(z/q)^{2}}$. 

The energy of an orbit is set by equating it  to the energy of a circular orbit in the plane of the disk at a radius of $R_{\rm circ}$.  In this work we focus on orbits with the energy of a circular orbit with $R_{\rm circ}=60$ kpc and $L_{z}=0.5L_{z,{\rm max}}$, which produces a range of orbits with diverse properties, and integrate 100 orbits in each library.
 
\begin{figure}
\centering
\epsscale{1.0}
\plotone{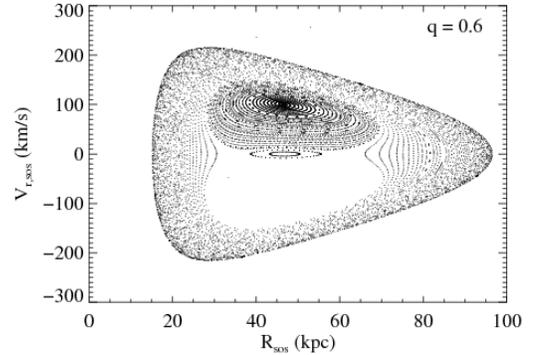}
\caption{\label{fig:sosplots} Surface of section for 100 orbits in a model with a dark matter halo with $q=0.6$.  All orbits have been integrated for at least 50 orbital periods. A large family of resonantly trapped orbits  is seen in a `resonant island', which appears as concentric ellipses centered on $(R_{\rm sos}, v_{R,{\rm sos}}) \simeq (45,100)$.}
\end{figure}

From these libraries we select orbits on which to initialize our progenitor using surface of section plots~\citep[e.g.,][]{binney_tremaine_87}.   Since resonant orbits conserve an additional integral of motion, their path collapses to a single point (or a small number of points, for higher-order resonances) in a surface of section.  Resonantly trapped orbits occupy a small region surrounding the parent orbit in the surface of section.  The surface of section for the orbits in our oblate halo model library is shown in Figure~\ref{fig:sosplots}.   The two concentric ellipses centered at $v_{R,{\rm sos}}=0$ in the surface of section are generated by orbits that are resonantly trapped with an orbit at the center of these ellipses.  This halo model also admits another resonant orbit family at roughly $R_{\rm sos}=45$ kpc, $v_{R,{\rm sos}}=100$ km/s. 

\begin{figure*}
\centering
\epsscale{0.8}
\plotone{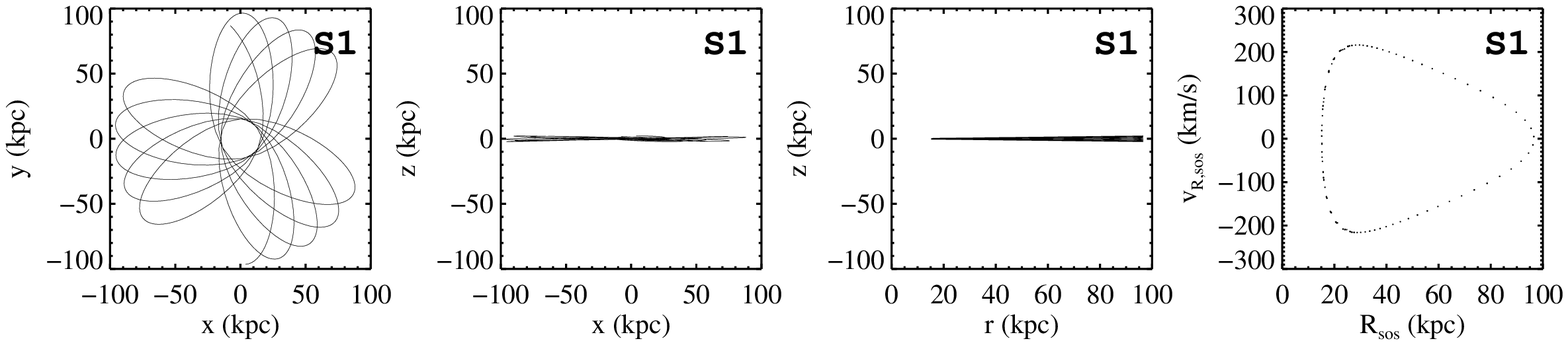}
\plotone{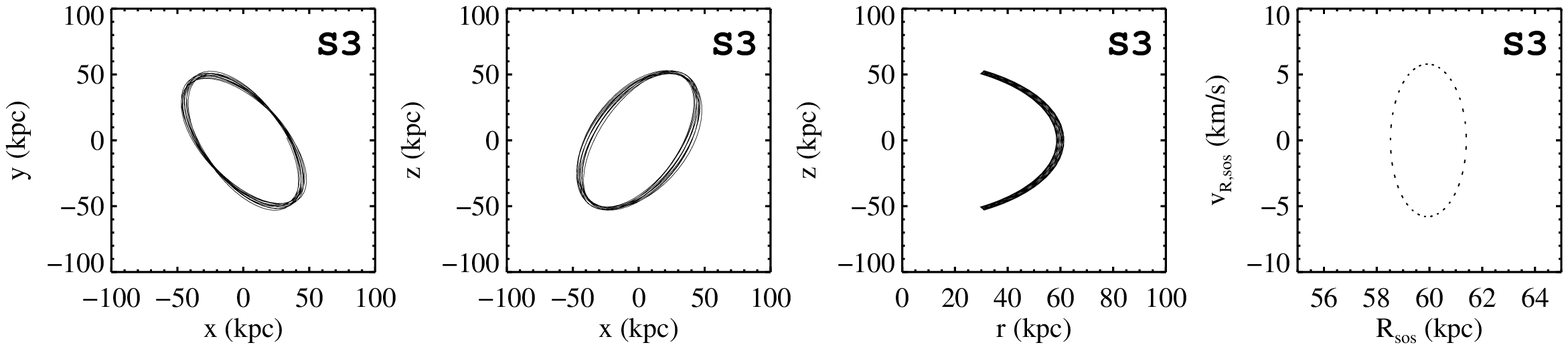}
\plotone{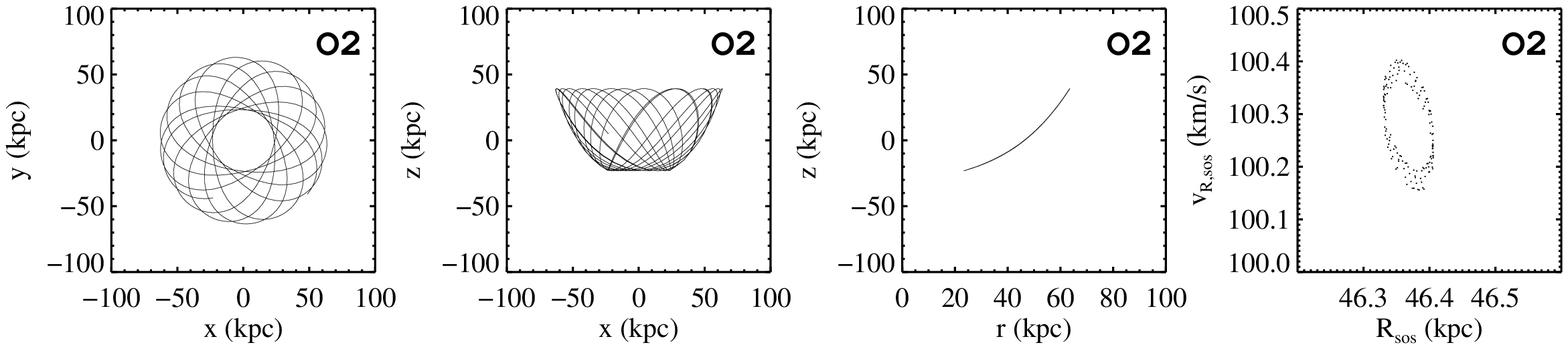}
\plotone{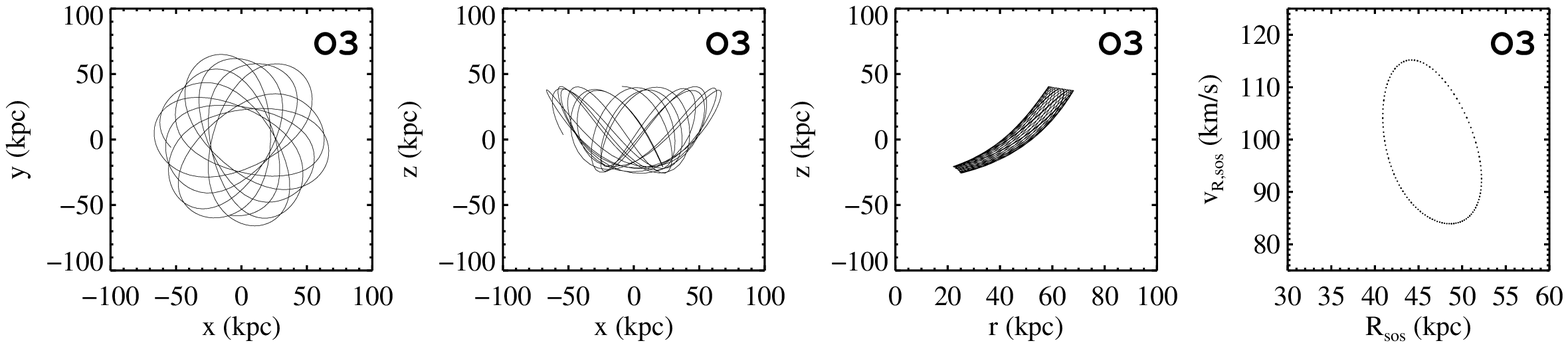}
\plotone{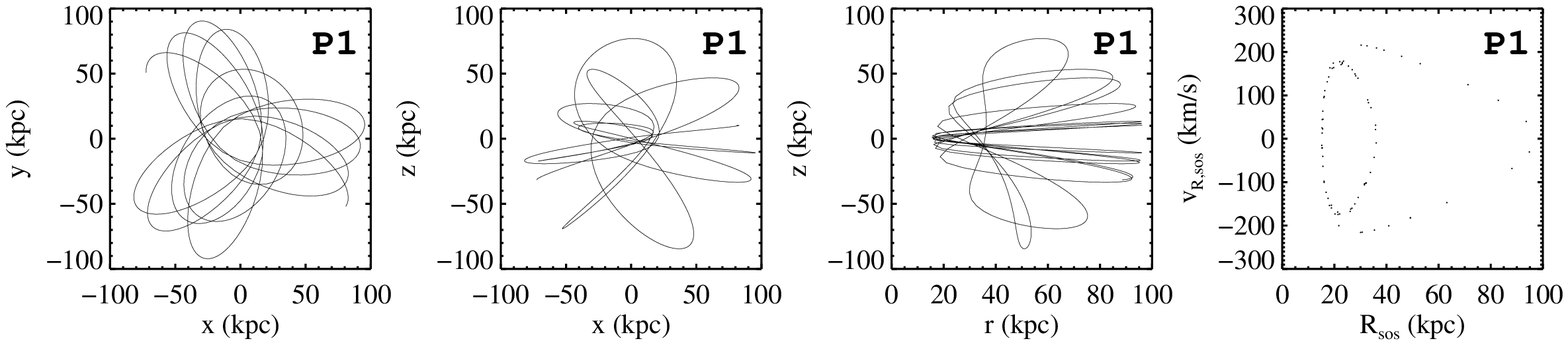}
\plotone{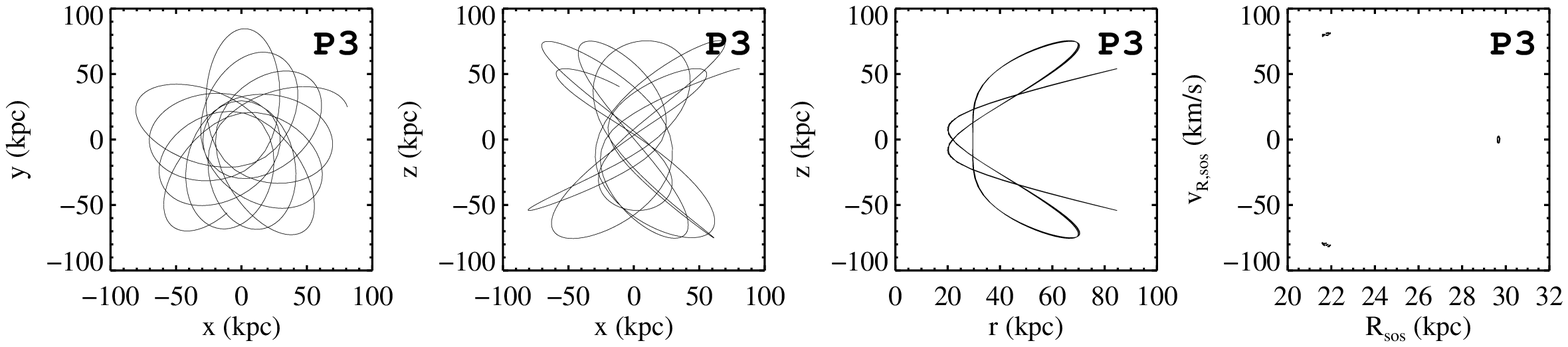}
\caption{Projections in $x$-$y$, $x$-$z$, $r$-$z$, and surface of section $R_{\rm sos}$-$v_{R,{\rm sos}}$ for selected orbits from each halo model.  S1 is a regular rosette orbit and appears as a nicely filled curve which covers a large range of radii and velocities in the SOS plot.  Orbit S3 is close to a resonance and occupies a much smaller region in the SOS plot (note that the axes have been rescaled).  O2 is extremely close to the strong resonance seen in the upper part of Figure~\ref{fig:sosplots}, and O3 is an orbit that lies at the edge of the resonant island.  Orbit P3 is close to a resonance and appears as three points in the SOS plot, while non-resonant orbit P1 fills a large area of the SOS plot.\label{fig:orbits}}
\end{figure*}

Properties of the orbits we have selected to simulate are given in Table~\ref{tbl:orbparams1}, with orbits which appear to be resonantly trapped labeled R in the surface of section (SOS) column, and non-resonant orbits labeled NR.  Also given are the apocenter ($r_{\rm apo}$) and pericenter ($r_{\rm peri}$) distances of each orbit.   
Projections of the orbital path and surface of section for some individual orbits are shown in Figure~\ref{fig:orbits}.  

While our method does not exactly identify any or all resonances in a given potential, we expect that the orbits we identify as resonantly trapped are sufficiently close to phase space resonances that they will exhibit the resonant behavior discussed in \S\ref{sec:phase}.    
Previous work has shown that relatively small perturbations do not destroy the strongest resonances \citep{sideris_kandrup_04}, therefore even though we choose orbital initial conditions from surfaces of section in smooth static potentials, we are confident that when we add substructure the resonantly trapped orbits continue to be 
associated with the same resonances, although their proximity to the resonance may change.

\begin{deluxetable}{ccccc}
\tablecaption{Orbit Parameters\label{tbl:orbparams1}}
\tablewidth{0pt}
\tablecolumns{5}
\tablehead{
\colhead{Halo}		&	\colhead{Orbit} 	& 	\colhead{$r_{\rm apo}$} 	&	\colhead{$r_{\rm peri}$}	& 	\colhead{SOS}\\
\colhead{$q$} 		&			&	\colhead{(kpc)}			&
	\colhead{(kpc)}			&	}
\startdata
0.6	&	O1	&	86.2	&	23.2	&	NR	\\
	&	O2	&	74.7	&	32.8	&	R	\\
	&	O3	&	77.8	&	30.1	&	NR	\\
	&	O4	&	54.2	&	43.2	&	R	\\
	&		&		&		&		\\
1.0	&	S1	&	96.6	&	15.2	&	NR	\\
	&	S2	&	74.2	&	44.9	&	NR	\\
	&	S3	&	61.4	&	58.5	&	R	\\
	&		&		&		&		\\
1.3	&	P1	&	99.0	&	15.2	&	NR	\\
	&	P2	&	102.1	&	18.7	&	NR	\\
	&	P3	&	101.7	&	20.6	&	R	\\
	&	P4	&	102.9	&	23.2	&	R	\\
\enddata
\end{deluxetable}

\section{Numerical Simulations}
\label{sec:sims}

\begin{deluxetable*}{ccccc}
\tablecaption{\label{tbl:subsoft}Subhalo Group Properties}
\tablewidth{0pt}
\tablecolumns{5}
\tablehead{
\colhead{Mass Range} & 		\colhead{Number of} 	 &  \colhead{$\epsilon_{\rm sub}$} & \colhead{$|a_{\rm soft}|_{\rm max}$ } & \colhead{$|a_{\rm NFW}|_{\rm max}$ }\\
\colhead{($10^{10} M_{\odot}$)} & 		\colhead{Subhalos}  & \colhead{(kpc)}  & \colhead{($10^{3}$ km/s/Gyr)} & \colhead{($10^{3}$ km/s/Gyr)}}
\startdata
4.62		&		1			&	6.64	&	3.03&	3.07\\
0.1 -- 1.16 &		12			&	3.57	&	2.61&	2.64\\
0.01 -- 0.1 &		67			&	1.25	&	1.85&	1.88\\	
0.0008 -- 0.01 & 	173			&	0.46	&	1.38&	1.40\\
\enddata
\end{deluxetable*}

We simulate the tidal disruption of our satellites using the N-body simulation code GADGET-2~\citep{springel_05}, which calculates gravitational forces using a hierarchical multipole expansion algorithm.  
The host galaxy model described in \S\ref{sec:select} is incorporated with a static potential using a modification to GADGET-2 kindly provided by B. Robertson.  

The initial satellite is modeled with $5 \times 10^{5}$ particles following a NFW density profile with virial mass $M_{\rm vir}=10^{10}$  M$_{\odot}$ and concentration parameter $c=12$.  
Our procedure for bringing the initial satellite into quasi-equilibrium with the host potential, described below, results in a progenitor of $\sim 3 \times 10^{9}$ M$_{\odot}$, roughly the mass of a dwarf galaxy such as Sgr.  

In this work we focus on a dwarf galaxy as our progenitor, as opposed to a globular cluster as in, e.g., \citet{ibata_lewis_irwin_quinn_02}, because we expect that a dwarf galaxy will produce debris which is more spatially extended due to its substantially greater mass and intrinsic velocity dispersion, and therefore will present a much larger cross section for interactions with substructures.  Since the most massive subhalos are rare and our simulations are evolved for only a few orbital periods, this increased cross section for interaction may easily be the difference between a massive substructure encountering the debris at least once or not at all.  

The mass of the NFW profile diverges as $r \rightarrow \infty$, so we truncate the satellite's density profile outside of the virial radius $r_{\rm vir}$ using an exponential cutoff as in \citet{kazantzidis_etal_04}.  
We set the radial scale which determines the sharpness of the cutoff $r_{\rm decay}=0.35 r_{\rm vir}$, which results in a total mass of $\sim 1.1 \times M_{\rm vir}$.
The satellite initial conditions are generated using the exact distribution function as detailed in \citet{kazantzidis_etal_04}, and we acknowledge S. Kazantzidis for providing us with code which implements this method.  The gravitational softening length for the satellite particles is set to $\epsilon_{\rm sat}=0.07$ kpc.

This initialization scheme sets up a satellite which is in equilibrium when isolated, but will not be in equilibrium when placed in the host potential.  As a result, the satellite will initially lose mass at an artificially high rate due to the sudden introduction of an external tidal field until it has relaxed into the host potential.  To minimize this effect, we create a progenitor in quasi-equilibrium with the host potential.  We place the initial satellite on circular orbits in the plane of the disk at 30, 40, and 50 kpc, and track the mass loss rate over time.  For all three cases there is an initial period of rapid mass loss, but the mass loss rate stabilizes after less than 0.2 orbital periods.  We generate a `remnant'  to use as our progenitor by selecting all particles still bound to the satellite after 0.2 orbital periods for the satellite on the 50 kpc orbit.  The closest distance at which we initialize any of our orbits is $\sim54$ kpc, so we choose the 50 kpc orbit for generating the remnant to temper the most intense initialization shock our progenitor would experience. The remant we use as our progenitor contains $\sim140$k particles and has a mass of $\sim3\times10^{9}$ M$_{\odot}$.

For our simulations in halos with substructure, we incorporate as softened point masses 253 subhalos from halo G$_{2}$ in the cosmological N-body simulation of \citet{kravtsov_gnedin_klypin_04} and thank A.~Kravtsov for making this data available to us.  The mass range of these subhalos is $\sim10^{7}$-$10^{10}$ M$_{\odot}$. 
Since our host potential is not identical to that of the host in the cosmological simulation (e.g., ours includes baryonic components), using the final positions and velocities of the subhalos in the cosmological simulation as our initial conditions will not produce identical orbits.  Instead, we maintain the final distribution of the subhalos in configuration space and in the $z$-component of angular momentum $L_{z}$: we initialize each subhalo in our simulations at its final spatial coordinates in the cosmological simulation, and transform its velocity by requiring that the current position is a turning point of the orbit ($v_{R}=v_{z}=0$) and that its $L_{z}$ is preserved.  

\citet{navarro_etal_97} showed that density profiles of CDM halos are well-fit by a NFW profile over a wide range of masses and \citet{kazantzidis_etal_04} found that dark matter subhalos retain their steep inner density profile even when tidally stripped.   Thus ideally we would model the subhalos in our simulation using a NFW density profile, but for numerical simplicity we model the subhalos as spline-softened point masses.  
The subhalo particles are sorted by mass to form four groups with different softenings.  The subhalo softening length $\epsilon_{\rm sub}$ for each group is chosen such that the maximum acceleration due to the most massive spline-softened subhalo $|a_{\rm soft}|_{\rm max}$ in each group  is equal to the NFW acceleration at $r=0.01r_{s}$ ~\citep[a radius well-within that shown to be stable against tidal disruption by][]{kazantzidis_etal_04}.   

The spline-softened acceleration is maximized at $r\sim0.8\epsilon$, while the NFW acceleration increases toward a finite limit as $r \rightarrow 0$, so using a spline softened potential increases the effective size of the subhalos.  We note that for the NFW profile the acceleration at $r=0.01 r_{s}$ is $\sim 99$\% of the maximum achieved by the NFW profile $|a_{\rm NFW}|_{\rm max}$.  However, large accelerations are reached even at moderate radii for a NFW profile: e.g., the NFW acceleration attains $\sim 40$\% of the maximum at $r=r_{s}$.  The functional form of the softened potential requires the softened acceleration to exceed the NFW acceleration for a small range of radii in order for the softened acceleration to anywhere reach the value for the NFW model at $r=0.01r_{s}$.  Requiring the softened acceleration to be smaller than the NFW acceleration at all radii would restrict the maximum softened acceleration to significantly less than the NFW maximum, so in order to more fully investigate the effect that the steep inner profile of NFW subhalos may have on tidal streams we allow the softened acceleration to exceed the NFW value at some radii. 

The softening for each group is set by first determining the concentration parameter $c$ for the most massive subhalo in the group using the $c(M_{\rm vir})$ relation determined by \citet{bullock_etal_01}, which gives $18\lesssim c \lesssim 40$ for our subhalos. The NFW scale radius $r_{s}$ is then determined from $c$ and $M_{\rm vir}$ for the most massive subhalo, and the softening length is then set from $r_{s}$ as described above.  Table~\ref{tbl:subsoft} summarizes the characteristics of each group.

We do not explicitly include a stellar component in our satellite model, but we attempt to follow the distribution of a stellar component by identifying a subset of dark matter particles as `star particles'.  In the initial equilibrium halo, star particles are tagged according to a Hernquist distribution.  After the remnant is generated there are $\sim$12,000 star particles remaining.

For each selected orbit, we simulate the disruption of the remnant progenitor in the host model both with and without dark matter subhalos, evolving our simulations for 4.9 Gyr.  

\section{Results}
\label{sec:results}

\subsection{Sky distribution of the debris}

\begin{figure*}[h]
\centering
\epsscale{1.0}
\plotone{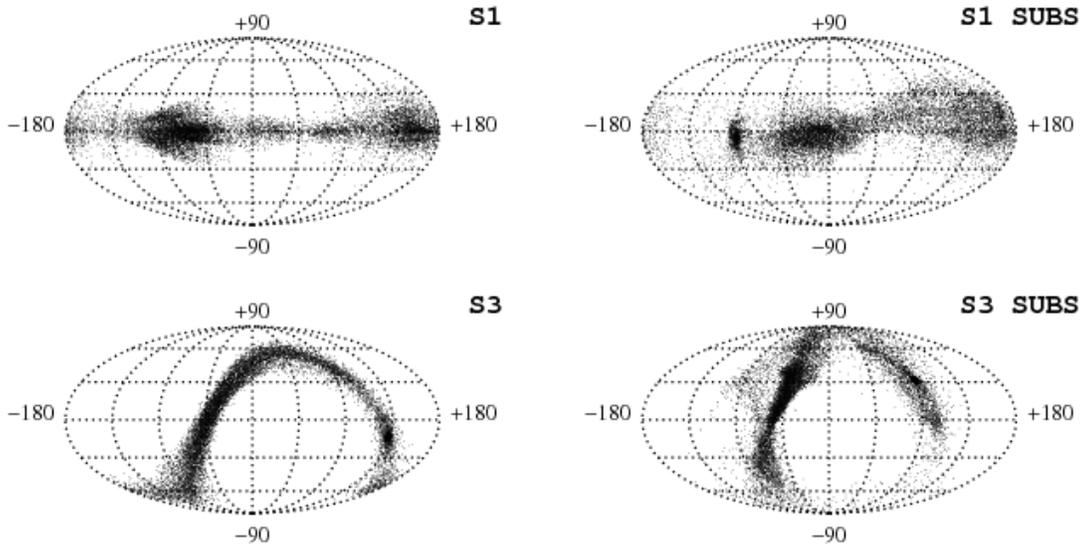}
\caption{Sky projection in Galactic coordinates of star particles after 4.9 Gyr for selected orbits in the spherical halo model.  Orbits in models with subhalos often produce clumpier debris than in models with a smooth halo component.  The additional disruption due to substructure is of a similar degree for the resonantly trapped (S3) and non-resonant (S1) orbits.  The S1 orbit with subhalos has clearly been perturbed, displacing a large amount of the debris relative to the smooth halo simulation. \label{fig:sph_lb_5gyr}}
\end{figure*}

\begin{figure*}[h]
\centering
\epsscale{1.0}
\plotone{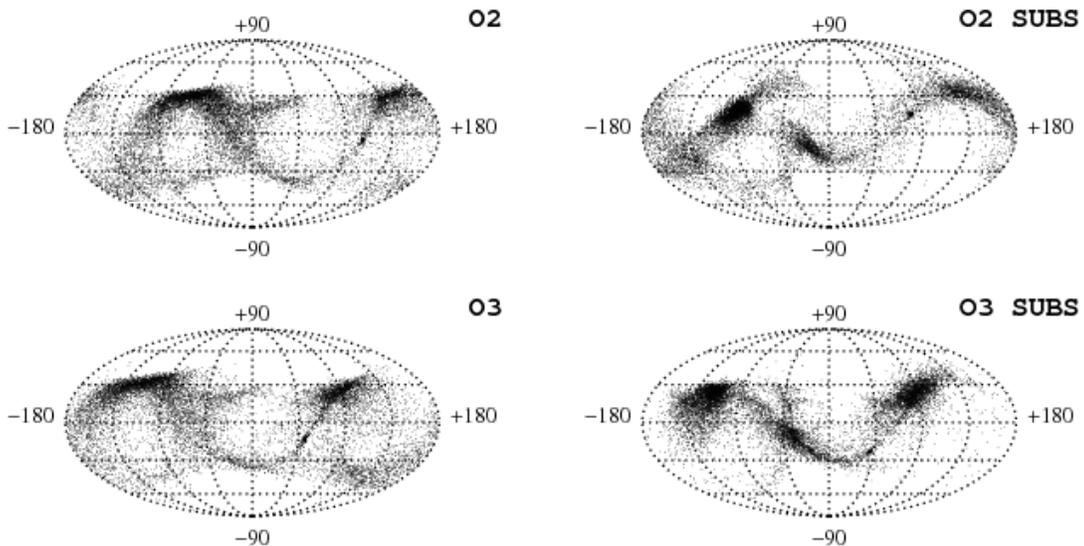}
\caption{Sky projection in Galactic coordinates of star particles after 4.9 Gyr for selected orbits in the oblate halo model.  The debris for orbits in models with subhalos appears slightly more clustered in this projection than for the models without subhalos, with most of the debris following a single path.  Both the resonantly trapped orbit (O2) and the non-resonant orbit (O3) produce debris that is fairly dispersed, regardless of the presence of subhalos.   The O3 simulation with subhalos produces an interesting feature at $\ell \sim-45^{\circ}$, $0^{\circ}\lesssim b \lesssim 30^{\circ}$ resembling a bifurcation in the stream.\label{fig:obl_lb_5gyr}}
\end{figure*}

\begin{figure*}[t!]
\centering
\epsscale{1.0}
\plotone{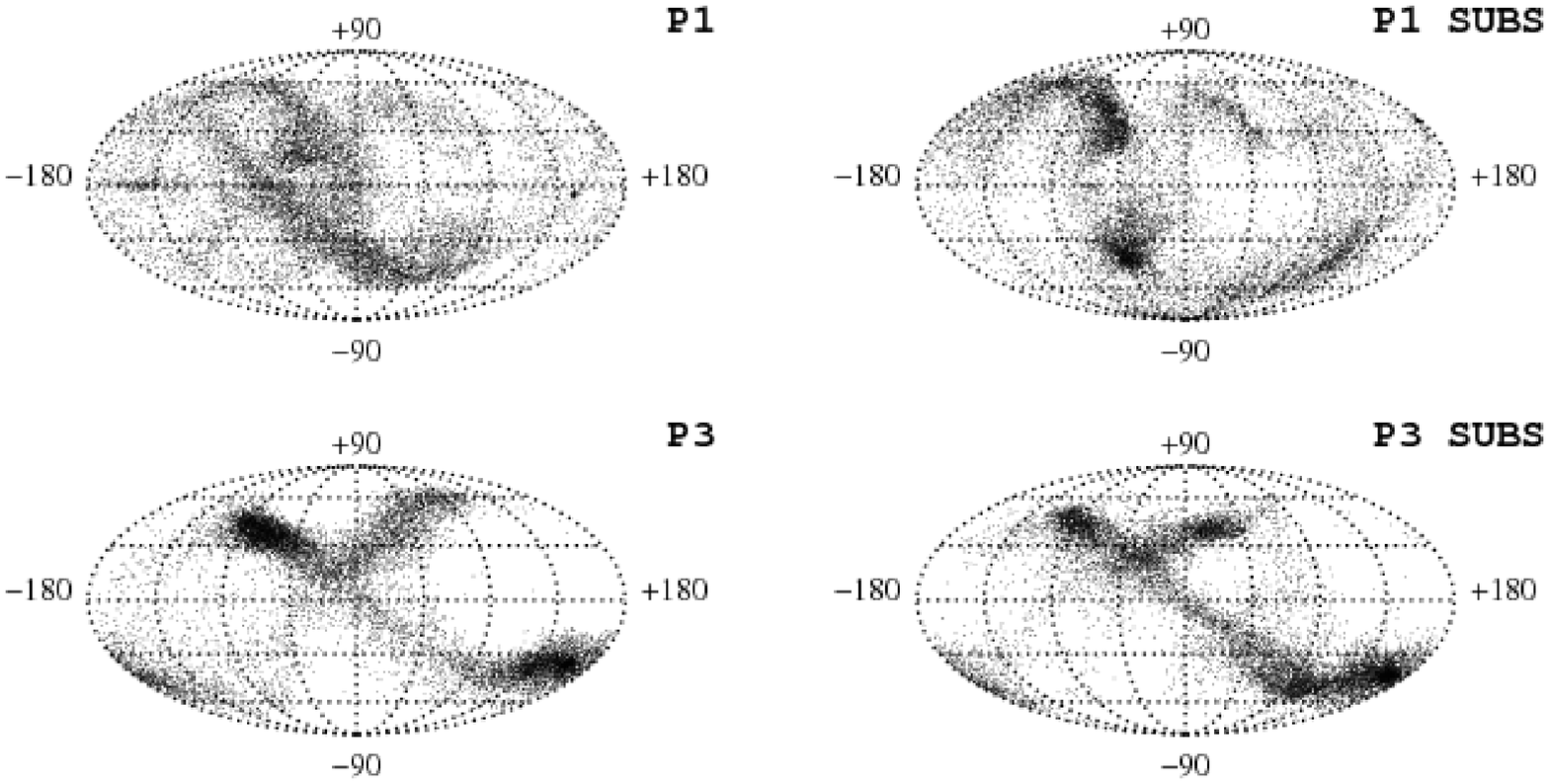}
\caption{Sky projection in Galactic coordinates of star particles after 4.9 Gyr for selected orbits in the prolate halo model.  The addition of  subhalos results in slightly more clustered debris in the case of P1, while the debris from orbit P3 qualitatively shows little difference between models with and without subhalos, producing streams in both cases.  \label{fig:pro_lb_5gyr}}
\end{figure*}

The projection of the star particles in Galactic coordinates after 4.9 Gyr of evolution for a subset of  orbits  is shown in Figures~\ref{fig:sph_lb_5gyr} (spherical halo model),~\ref{fig:obl_lb_5gyr} (oblate halo model), and~\ref{fig:pro_lb_5gyr} (prolate halo model).  Comparing the simulations in all three halo models, it is immediately notable that even those without subhalos do not consistently produce thin, coherent streams.  Simulations with subhalos often lead to clumpier structures in these coordinates, but the degree of additional scattering or clumping for the same orbit simulated with and without subhalos is small compared to the variation seen between different orbits and different halo models without subhalos.  

In Galactic coordinates, the S3 orbit without subhalos produces the most coherent stream, which is not surprising given that this orbit has $r_{\rm apo} \sim r_{\rm peri} \sim 60$ kpc.  The S1 orbit is largely confined to the plane of the disk of the Galaxy and has a large apocenter to pericenter ratio ($r_{\rm apo}/r_{\rm peri} \gtrsim 6$), and in this projection produces clumpy debris clustered along the Galactic plane in the smooth halo model.  For both orbits in the spherical potential the presence of substructure produces some additional dispersion.  Substructure also leads to noticeably clumpier debris for the S3 orbit, and results in debris displaced from the Galactic plane for orbit S1.  

In the oblate and prolate halo models, the resulting streams are quite dispersed compared to the S3 stream and somewhat clumpy both in the simulations with and without subhalos.  For the orbits shown in the oblate halo, O2 and O3, as well as the P1 orbit in the prolate halo, the debris appears more clustered in the simulations with subhalos.  Orbit P3 produces somewhat smooth streams which are qualitatively very similar both with and without subhalos, suggesting it has not had any significant interactions with the subhalo population.

Resonantly trapped orbits S3 and P3 produce slightly more coherent streams than the non-resonant orbits in the same halos.  Yet, both orbit O2, which is closely associated with a resonance, and the non-resonant orbit O3 produce qualitatively similar debris which is fairly dispersed.  
In this limited sample of orbits we do not find that resonantly trapped orbits consistently produce substantially more coherent streams than non-resonant orbits, and it does not appear that these orbits are less susceptible to the effects of substructure than the non-resonant orbits.

In many cases, orbits in models with subhalos result in sections of debris in very different locations than in the smooth halo models.  The S1 orbit with subhalos produces a stream that is clearly offset from the Galactic plane for $\ell \gtrsim0^{\circ}$, while the stream from the same orbit in the model without substructure wraps around the Galaxy at $b\sim0^{\circ}$.  Similar formations can be seen in most of the simulations with substructure, suggesting that disturbances in tidal streams due to interactions with substructures (both dark and luminous) in the halo may be an important consideration when constructing models to fit observations of tidal streams.  Of particular interest is a small stream of stars slightly offset from the main stream at $\ell \sim -45^{\circ}, 0^{\circ} \lesssim b \lesssim 30^{\circ}$ in the O3 simulation with subhalos which in these coordinates qualitatively resembles the bifurcation seen in the Sgr stream \citep{fellhauer_etal_06}.  This feature is not apparent in the smooth halo simulation.  We do not in any way suggest that our orbit O3 is an appropriate description of the orbit of Sagittarius, but draw attention to this feature as an example of the type of effect substructure can have on tidal streams which may complicate the interpretation of observed tidal streams.

\subsection{Structure in phase space coordinates}

\begin{figure*}[t!]
\centering
\epsscale{0.8}
\plotone{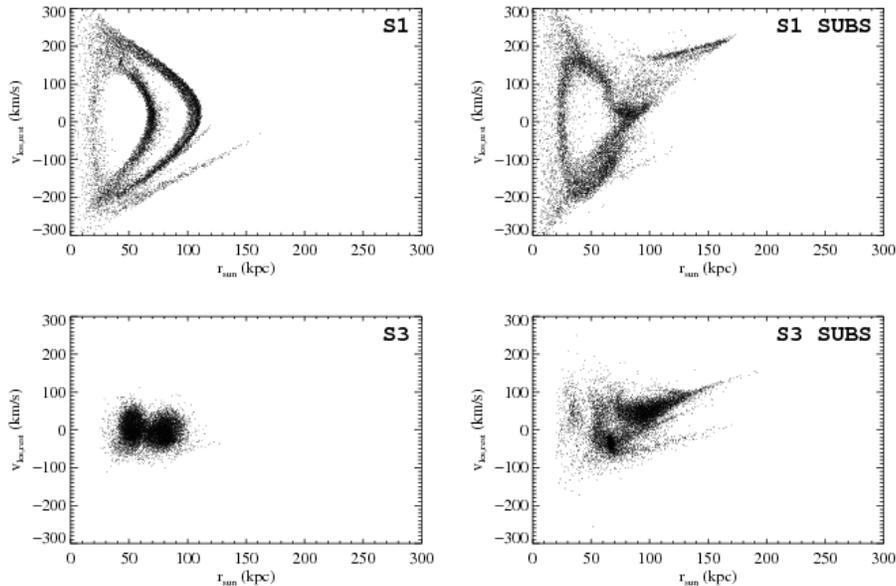}
\caption{Heliocentric line-of-sight velocity in the rest frame of the Galaxy versus heliocentric distance of star particles in the spherical halo model after 4.9 Gyr for the same orbits as in Figure~\ref{fig:sph_lb_5gyr}.  
Simulations with subhalos produce arms of debris in these coordinates at relatively large radii ($r_{\rm sun} \gtrsim 150$ kpc) strongly correlated in velocity.  This feature is also seen for many orbits in oblate and prolate models with subhalos.  The S3 simulation with subhalos produces multiple arms of debris, the longest of which extends to $\sim200$ kpc.  For comparison, the debris from the same orbit in the simulation without subhalos scarcely extends beyond $\sim100$ kpc.\label{fig:sph_rsun_vrlos}}
\end{figure*}

\begin{figure*}[t!]
\centering
\epsscale{0.8}
\plotone{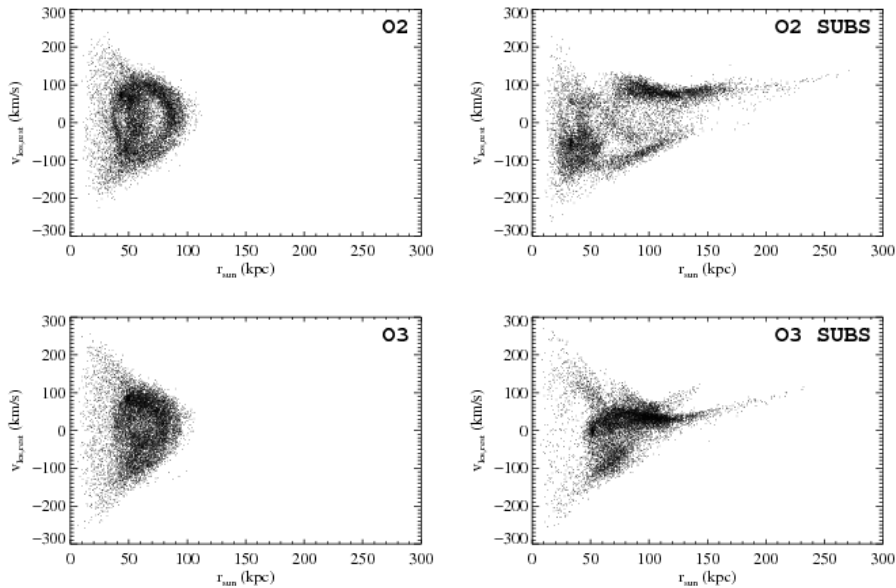}
\caption{Heliocentric line-of-sight velocity in the rest frame of the Galaxy versus heliocentric distance of star particles in the oblate halo model after 4.9 Gyr for the same orbits as in Figure~\ref{fig:obl_lb_5gyr}.  As in the other halo models, arms of star particles at large radii are present for orbits in simulations with subhalos, with the O2 arm stretching past 250 kpc.  Neither of these orbits shows significant structure in these coordinates when simulated in a smooth halo, even though O2 is a resonantly trapped orbit.\label{fig:obl_rsun_vrlos}}
\end{figure*}

\begin{figure*}[t!]
\centering
\epsscale{0.8}
\plotone{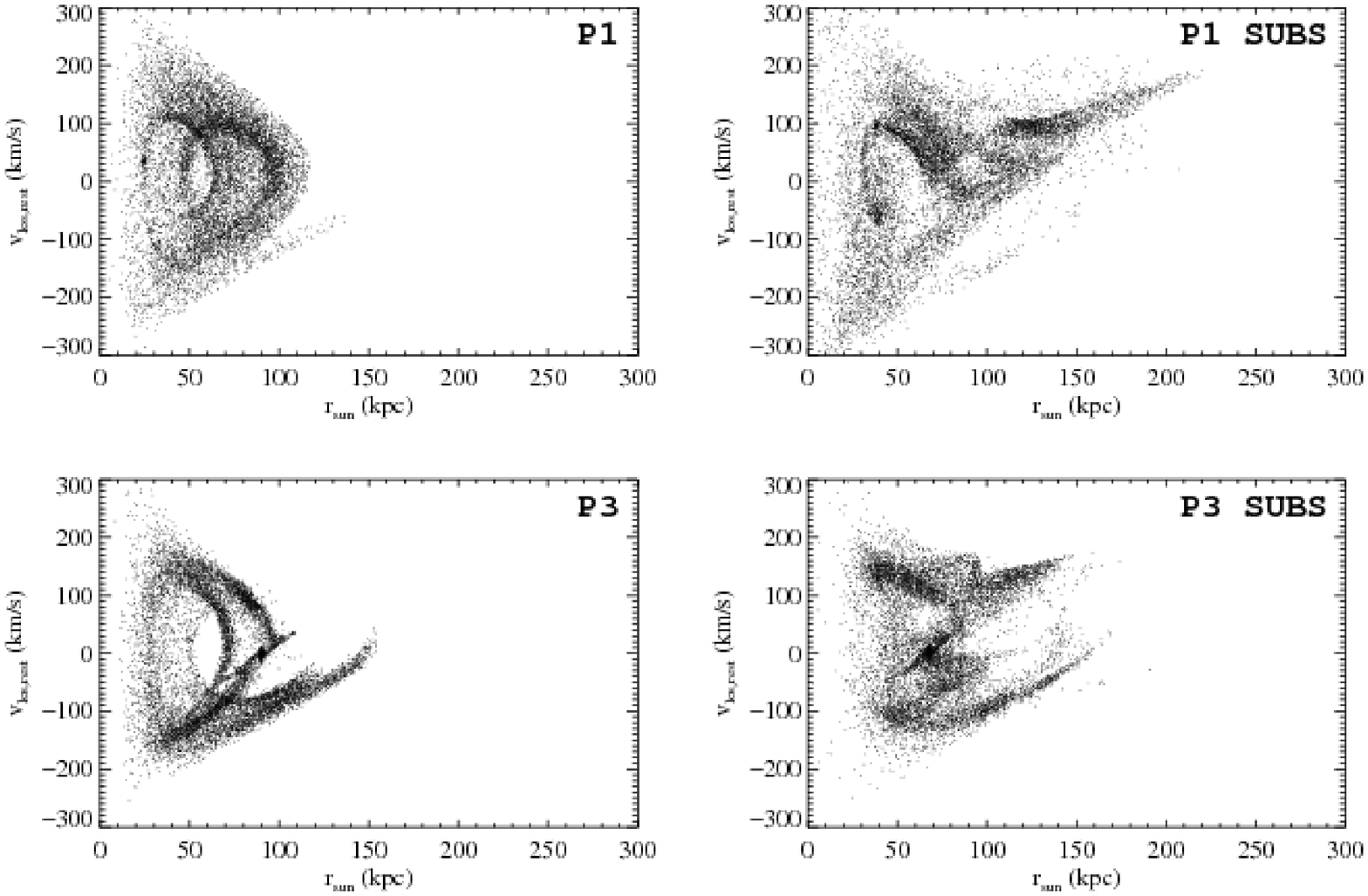}
\caption{Heliocentric line-of-sight velocity in the rest frame of the Galaxy versus heliocentric distance of star particles in the prolate halo model after 4.9 Gyr for the same orbits as in Figure~\ref{fig:pro_lb_5gyr}.  Like the simulations in oblate and spherical halo models with subhalos, the P1 simulation with subhalos produces debris at large radii correlated in velocity, but this feature is not seen as clearly in the P3 simulation with subhalos.\label{fig:pro_rsun_vrlos}}
\end{figure*}

The distribution of the star particles in line-of-sight velocity in the rest frame of the Galaxy $v_{\rm los,rest}$ and heliocentric distance $r_{\rm sun}$ is shown in Figures~\ref{fig:sph_rsun_vrlos},~\ref{fig:obl_rsun_vrlos}, and~\ref{fig:pro_rsun_vrlos}.  
It is notable that simulations with subhalos tend to produce debris which is less coherent in these coordinates than debris in their corresponding simulations in the smooth halo models.  However, the variation in the patterns formed in the absence of substructure is quite large compared to the changes induced by the addition of substructure, as was the case for the sky distribution of the debris.  For example, orbits S1 and S3 in the smooth halo models produce drastically different patterns: the S1 simulation produces clear bands of debris in these coordinates, while the S3 simulation appears as two compact clumps with little internal structure.  

Many simulations with subhalos are distinguishable in these coordinates by a conspicuous trail of debris at large radii strongly correlated in $v_{\rm los,rest}$.  Furthermore, $v_{\rm los,rest}$ for the particles in this feature is {\it positive} and {\it increases} with increasing radius, unlike patterns formed by the wrapping of debris in smooth halo models (e.g., in the S1 and P3 orbits without subhalos).   
This feature is present to some degree in most of our simulations with subhalos in all halo shapes, and is consistently absent when the satellite is evolved in a smooth halo. 
These `arms' of star particles typically extend far beyond the debris seen in the smooth halo simulations.  Many orbits in simulations with substructure produce debris extending to $\sim200$ kpc, while debris from these orbits in smooth halo simulations is typically confined to $\sim100$ kpc.  
The particles in these arms are generally not very spatially clustered, so their strong correlation in velocity may prove essential for identifying stars as originating from the same progenitor, as well as for cleanly identifying this feature as the result of substructure.  We emphasize that this observable signature appears to be unique to simulations with subhalos, making a detection of such correlated stars a potential smoking gun indicating the presence of substructure in the halo, regardless of the shape of the dark matter halo.

It is important to note that some of the dark matter subhalos in our simulation may host a luminous component, making them analogous to known dwarf galaxies or other satellites of the Milky Way, so this signature could be evidence of interactions with luminous substructures as well as dark ones.  However, \citet{strigari_bullock_kaplinghat_etal_07} propose that the well-known luminous satellites of the Milky Way do not all inhabit the most massive dark subhalos, which in turn implies that some dark subhalos have masses comparable to the masses of the known luminous satellites.  Under this assumption, any signatures of substructure detected would not necessarily be more likely due to luminous structures than to dark ones.

To confirm that this striking feature does not result from lack of precision of the integration scheme, we repeated the S3 simulation with smaller timesteps to prevent unphysical two-body interactions, and found the same feature present.  Additionally, we repeated a number of simulations using the subhalos taken from host G3 in \citet{kravtsov_gnedin_klypin_04}, and again found that many of the simulations developed similar trails of stars at large radii, suggesting that the feature does not arise from a rare, serendipitous arrangement of the subhalos. 

\begin{figure}[t]
\centering
\epsscale{1.0}
\plotone{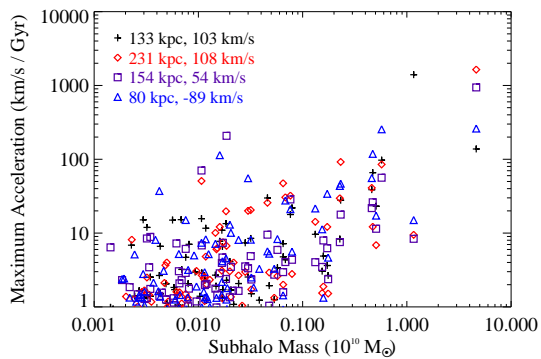}
\caption{({\it color online}) Maximum acceleration due to each subhalo experienced by four selected star particles versus subhalo mass for the O3 simulation.  Particles shown are selected from features seen in Figure~\ref{fig:obl_rsun_vrlos}: the shorter, upper arm ({\it black crosses}), the tip of the long arm ({\it red diamonds}), the middle of the long arm ({\it purple squares}), and the main region ({\it blue triangles}).  Final $r_{\rm sun}$ and $v_{\rm los,rest}$, as plotted in Figure~\ref{fig:obl_rsun_vrlos}, are labeled for each particle. \label{fig:max_acc}}
\end{figure}

To identify the mechanism leading to this feature, we examine the acceleration produced by the subhalos on selected star particles.  We estimate the maximum acceleration produced by a subhalo on a given star particle over the course of the simulation by determining the closest approach of the particle to the subhalo in any snapshot.  Due to the limited time resolution of our snapshots, the distance of closest approach determined in this way is likely to be significantly overestimated, and hence the maximum acceleration may be significantly underestimated.  

Figure~\ref{fig:max_acc} shows the maximum acceleration due to each subhalo as a function of subhalo mass for four selected star particles in the O3 simulation with substructure.  Each of the four particles was selected from a different region of the $r_{\rm sun}$-$v_{\rm los,rest}$ plot in Figure~\ref{fig:obl_rsun_vrlos}: a particle from the shorter, upper arm ({\it black crosses}), the tip of the longer arm ({\it red diamonds}), the middle of the longer arm ({\it purple squares}), and the lower part of the main region ({\it blue triangles}).  A clear trend is evident, with the most massive subhalos typically producing the largest accelerations.  A large acceleration does not guarantee a substantial energy exchange, but indicates the strength of an individual encounter.  All three particles from the arms had a strong encounter with a very massive subhalo ($\gtrsim 10^{10}$ M$_{\odot}$), while the particle not located in an arm (marked with a blue triangle) did not have any encounters producing accelerations greater than $\sim 300$ km/s /Gyr recorded in a snapshot.

\begin{figure*}
\centering
\epsscale{0.8}
\plottwo{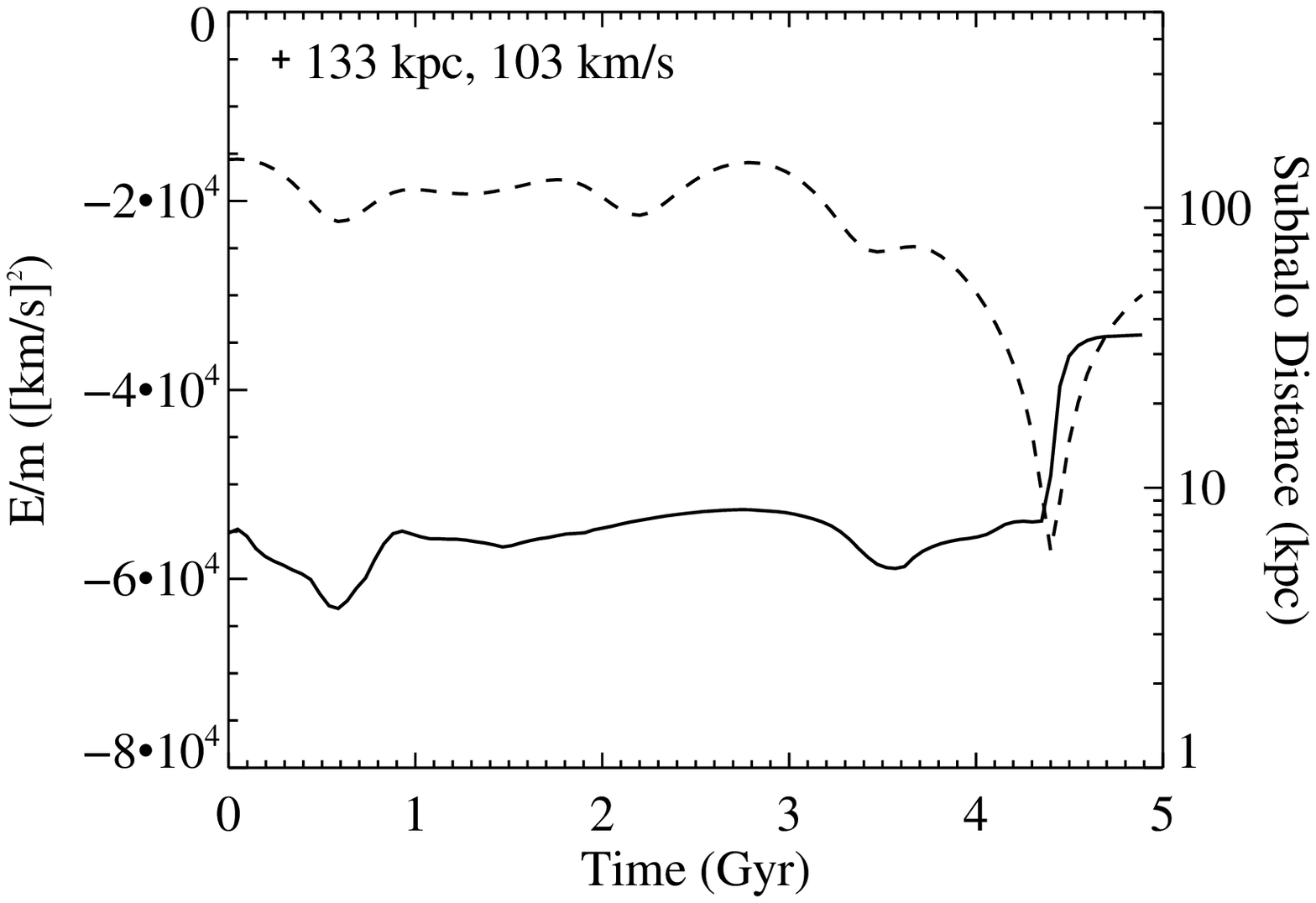}{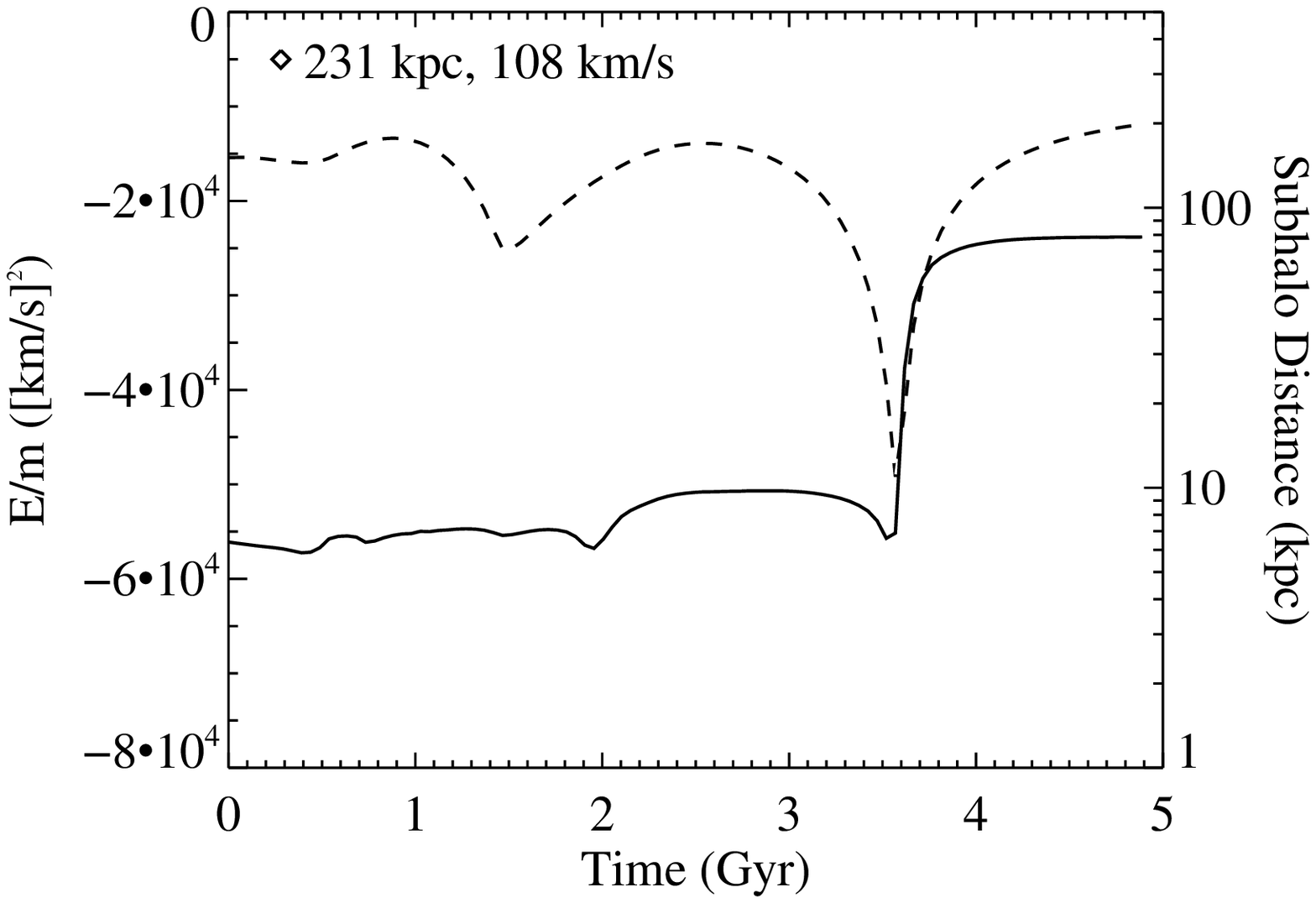}
\plottwo{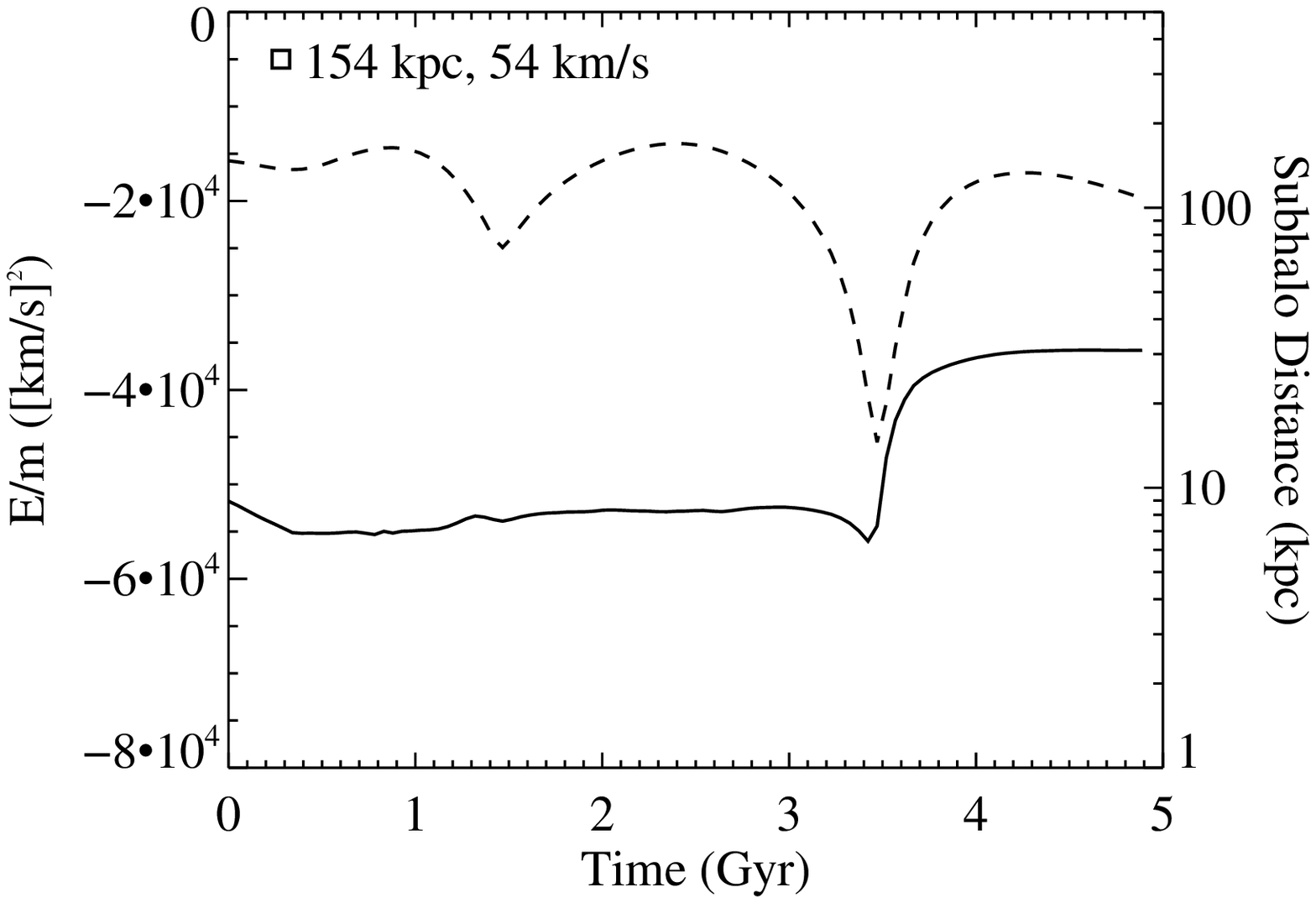}{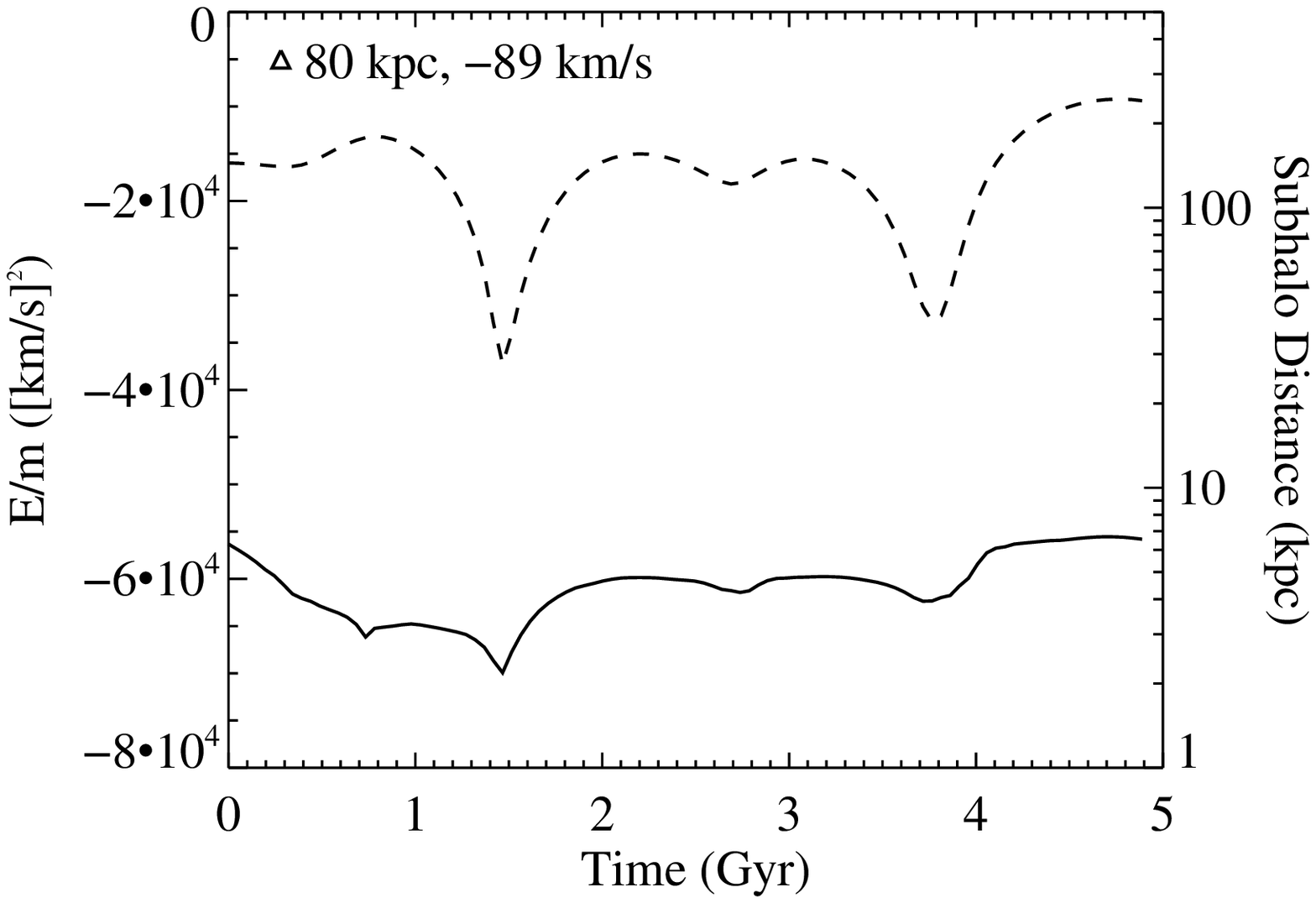}
\caption{Total energy per unit mass ({\it solid line}) and distance to subhalo producing the largest acceleration ({\it dashed line}) for four selected star particles in the O3 simulation.  Particles shown are the same as in Figure~\ref{fig:max_acc}.  Final $r_{\rm sun}$ and $v_{\rm los,rest}$, as plotted in Figure~\ref{fig:obl_rsun_vrlos}, are labeled for each particle.  The dashed line corresponds to the second most massive subhalo in the top left panel, and corresponds to the most massive subhalo in the other three panels.  A clear correlation is seen between large increases in energy and the proximity of the subhalo. \label{fig:etot_t}}
\end{figure*}

The time evolution of the total energy of selected particles is shown in Figure~\ref{fig:etot_t}, along with the proximity of each particle to the subhalo which caused it the greatest acceleration.  The particles selected correspond to those in Figure~\ref{fig:max_acc}.  For the three particles in the arms at the end of the simulation (the two top panels and the bottom left panel), a large increase in energy is evident, clearly corresponding to the time of closest approach of the subhalo.  The energy of the particle in the bottom right panel, which had a lesser maximum acceleration in Figure~\ref{fig:max_acc}, shows smaller deviations, but which still appear to be correlated with the proximity of the subhalo.  The net change in total energy for this particle over the course of the simulation is small compared to the increases in the other particles' energies.
For the three particles which did gain a substantial amount of energy, the majority of the increase occurred during a single interaction with the subhalo producing the largest recorded acceleration.  

\begin{figure*}[t]
\centering
\plottwo{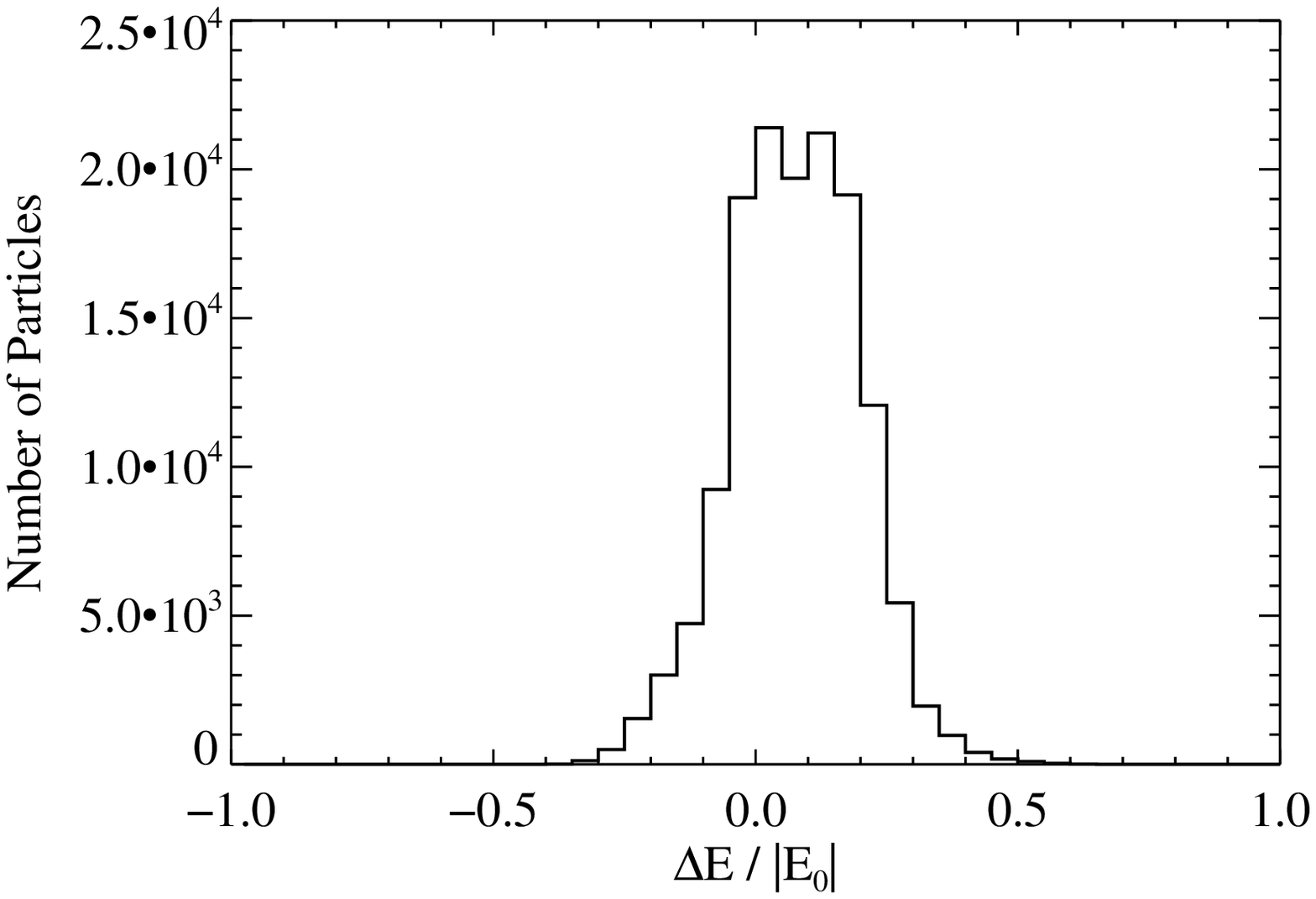}{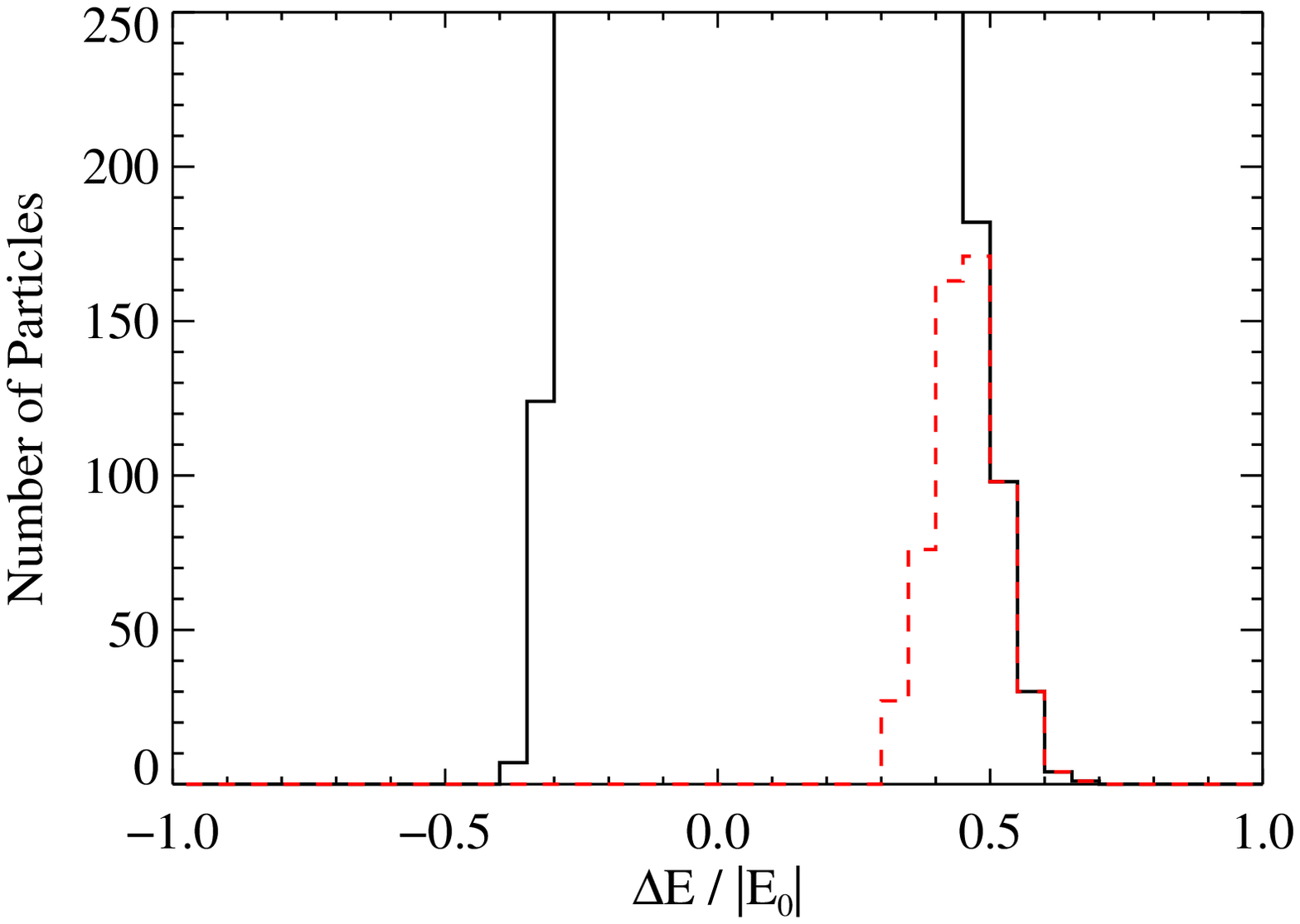}
\caption{Fractional change in total particle energy for all particles in the O3 simulation with substructure over the course of the simulation ({\it solid line}) and for all particles in the longer arm at the end of the simulation ({\it red dashed line, right panel, color online}).  Particles defined to be in the arm have $r_{\rm sun}>170$ kpc, and include dark matter particles and star particles.  $E_{0}$ is each particle's total energy at the beginning of the simulation.  The particles in the arm occupy the upper end of this distribution, accounting for the majority of particles with an increase in energy of at least $\sim 45$\%.\label{fig:frac_e_hist}}
\end{figure*}

Figure~\ref{fig:frac_e_hist} shows the distribution of the fractional change in total energy over the course of the simulation for all satellite particles (both dark matter and star particles) in the O3 simulation with subhalos.  The solid line shows the change in energy for all particles, and the red dashed line represents only the particles in the longer arm at the end of the simulation ($r_{\rm sun}>170$ kpc).  The small positive offset from zero seen for the distribution for all particles shows that satellite particles on average gained a small amount of energy ($\sim 8$\%), indicating that interactions with the subhalo population have a slight tendency to increase the energy of the stream particles.  The right panel is rescaled to show the distribution of the particles in the arm in relation to all the satellite particles.  The arm particles appear to naturally fill the positive end of the $\Delta E/|E_{0}|$ distribution, accounting for almost all of the particles which experienced an energy increase of $\sim 45$\% or more.  In addition, all of the 55 star particles in this feature experienced an acceleration from a single subhalo of at least 600 km/s /Gyr, and for all but one of these particles the most massive subhalo was responsible for the strongest recorded encounter.  

A number of mechanisms for transferring large amounts of energy to stream particles are possible, including two-body encounters (i.e., the acceleration and energy gain of the particle is due primarily to the influence of a single subhalo), three-body encounters (i.e., the energy gain is due to two subhalos producing comparable accelerations on the particle during the time the majority of the energy transfer occurs), and less easily categorized interactions (e.g., a particle-subhalo encounter during which the acceleration of the subhalo by the host potential provides a non-negligible contribution to the particle's energy gain). 
We have examined in detail the circumstances of the large energy gain for a few selected particles, deferring a thorough investigation of the energy exchange mechanism to a future study.
In our simulations we find examples of an isolated two-body interaction between a particle and a subhalo and of a particle-subhalo interaction during which the acceleration of the host potential appears to play an important role.  We do not find clear evidence of three-body interactions, but our limited inquiry does not rule out the possibility of such interactions.

To test the sensitivity of our results to the subhalo softening scheme, we consider the change in a particle's total velocity by an isolated interaction with a massive subhalo modeled by a NFW or a softened potential.
The angular deflection of a particle's trajectory by a central force can be determined by integrating the orbit equation:
\begin{equation}
\label{eqn:scattangle}
\Theta = \pi - 2\int_{r_{\rm min}}^{\infty} \frac{\ell \ dr}{ r^{2} \sqrt{ 2(E - U(r) - \frac{\ell^2}{2 r^{2}}) } }
\end{equation}
where $E$ and $\ell$ are the total energy and angular momentum of the particle per unit mass, $U(r)$ is the potential per unit mass, and $r_{\rm min}$ is the distance of closest approach of the particle.  Defining the impact parameter $b$ to be the perpendicular distance between the undeflected path of the particle and the center of force, $\ell = b \sqrt{2 E} = b v_{\rm rel}$ with $v_{\rm rel}$ the relative velocity before the encounter.  The scattering of a particle by a substantially more massive subhalo is well-described by this setup as long as the subhalo does not accelerate significantly during the interaction.

\begin{figure}[t]
\centering
\epsscale{1.0}
\plotone{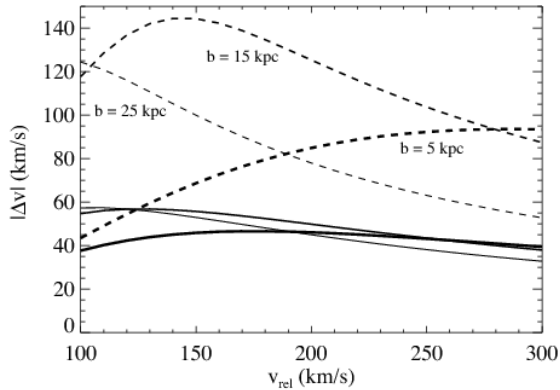}
\caption{Change in velocity $|\Delta \vec{v}\,| = |\vec{v_{f}}-\vec{v_{i}}|$ for a particle scattered by the NFW ({\it solid}) and softened ({\it dashed}) potentials for the most massive subhalo in our simulations.  The thick, medium, and thin lines are for $b = $ 5, 15, and 25 kpc respectively.\label{fig:deltav}}
\end{figure}

The magnitude of the change in a particle's velocity during this interaction $|\Delta \vec{v}\,| = |\vec{v_{f}}-\vec{v_{i}}|$ is determined for a given $b$ and $v_{\rm rel}$.  Figure~\ref{fig:deltav} shows $|\Delta \vec{v}\,|$ as a function of $v_{\rm rel}$ for selected values of $b$ using the NFW and the softened potentials for the most massive subhalo in our simulations.  For an isolated scattering interaction with this subhalo, the maximum $|\Delta \vec{v}\,|$ is $\sim 60$ km/s for the NFW potential (at $v_{\rm rel} \sim 110$ km/s, $b \sim 22$ kpc), with a fairly weak dependence on $v_{\rm rel}$ and $b$.
For the softened potential, $|\Delta \vec{v}\,|$ is more strongly dependent on $v_{\rm rel}$ and $b$, achieving a maximum of $\sim 145$ km/s (at $v_{\rm rel} \sim 130$ km/s, $b \sim 14$ kpc), with a typical $|\Delta \vec{v}\,|$ somewhat smaller depending on the parameters.  $|\Delta \vec{v}\,|$ reaches a minimum of $\sim 35$ km/s for the NFW potential and $\sim 43$ km/s for the softened potential for $5 < b < 20$ kpc and $100 < v_{\rm rel} < 300$ km/s. 

This calculation suggests that our choice of softening may lead to changes in the particle's velocity larger by up to a factor of $\sim 2.5$ for the strongest isolated interactions than if we had implemented the NFW potential.  However, we emphasize that this calculation only gives an estimate of the change in a particle's velocity, which cannot be easily translated to an effect on the properties of the debris without detailed information about the parameters of a particular interaction in the rest frame of the Galaxy.  The change in the particle's kinetic energy in the rest frame of the Galaxy depends on the initial velocity vector of the particle in the Galactic rest frame, and so these scattering events can reduce as well as increase the total energy of the particles.  

From this analytical estimate, we expect that any features in our simulations attributable to interactions with subhalos, such as the arms of debris in Figures~\ref{fig:sph_rsun_vrlos},~\ref{fig:obl_rsun_vrlos}, and~\ref{fig:pro_rsun_vrlos}, could be somewhat less prominent for subhalos modeled with a NFW potential.  
Because the parameters determining the energy gain or loss of a particle may vary greatly between scenarios in which scattering events occur, it is not possible to determine, in general, how much less prominent these features will be for NFW subhalos, or how often the resulting debris will differ substantially from that obtained using softened subhalos.  

Interactions in which the acceleration of the subhalo by the host galaxy contributes appreciably to the energy gain of the particle are also seen in our simulations.  In these cases the two-body scattering approximation is not an adequate description of the interaction, and our use of a spline-softened instead of a NFW potential for the subhalo plays a smaller role in determining the energy gain of the particle.

\begin{figure*}[th]
\centering
\epsscale{0.8}
\plotone{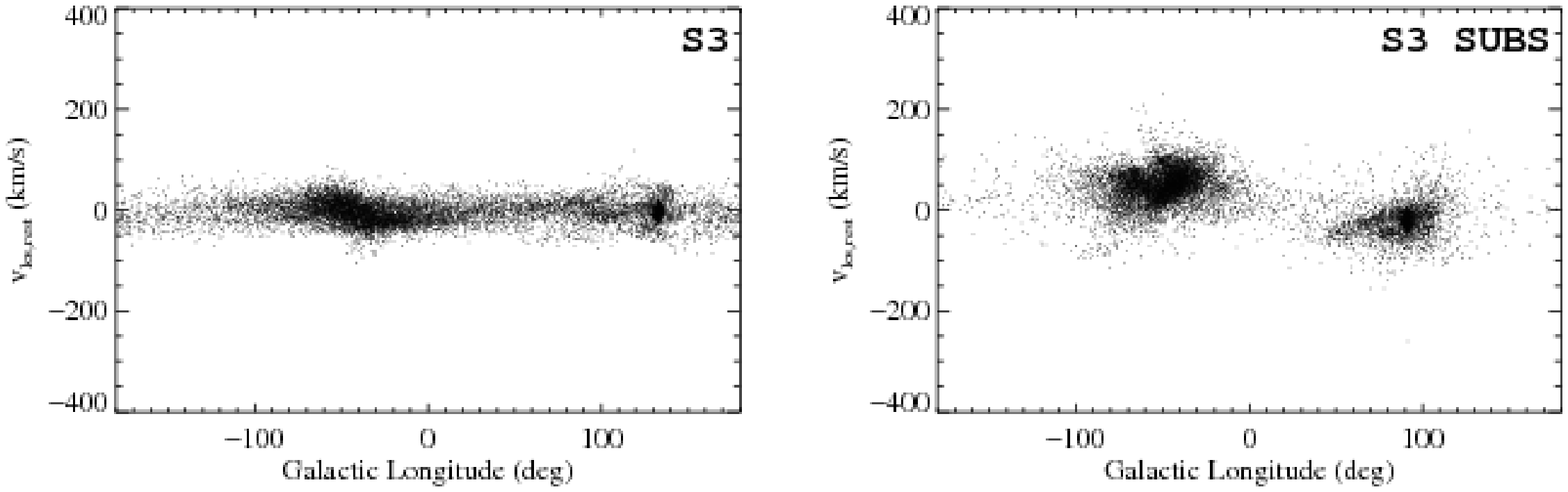}
\plotone{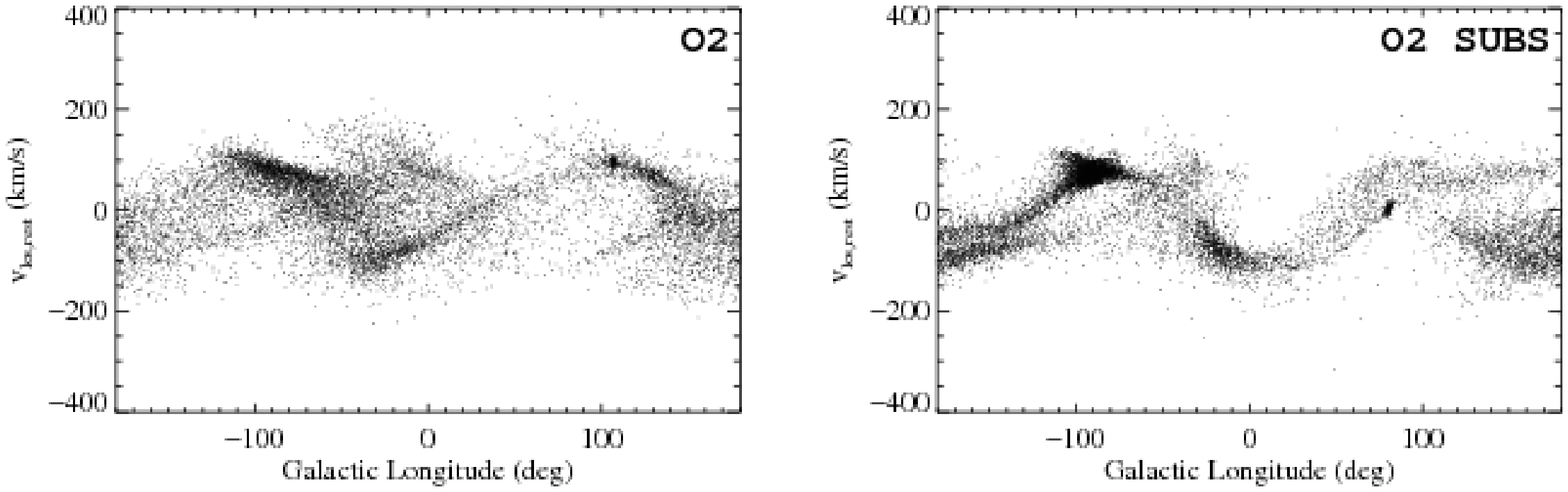}
\plotone{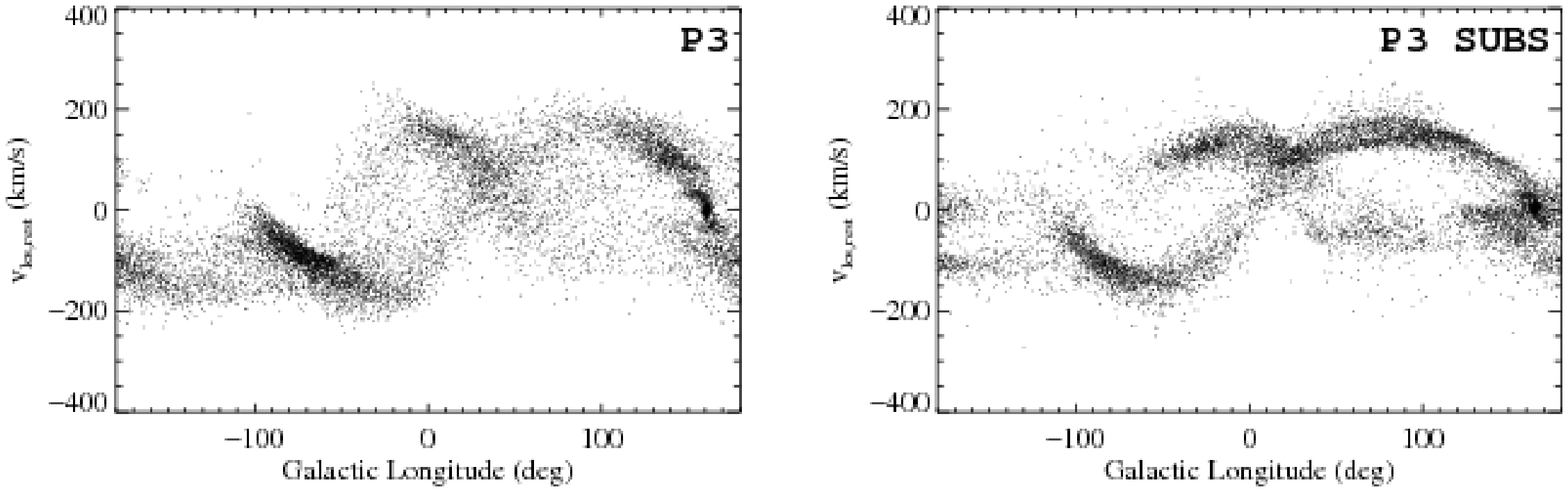}
\caption{Line-of-sight velocity in the rest frame of the Galaxy versus Galactic longitude of star particles after 4.9 Gyr for S3, O2, and P3 simulations with and without substructure.  Contrary to expectations, narrower streams in these coordinates are produced in the presence of substructure than in the smooth halo for orbits O2 and P3.   \label{fig:l_vrlos}}
\end{figure*}

Previous work \citep{ibata_lewis_irwin_quinn_02, johnston_spergel_haydn_02, mayer_etal_02, penarrubia_etal_06} has suggested that interactions with substructure would increase the velocity dispersion of the streams.  To determine whether the presence of substructure can be inferred from this measurement, we examine the spread in $v_{\rm los,rest}$ as a function of Galactic longitude for star particles after 4.9 Gyr, shown in Figure~\ref{fig:l_vrlos} for three selected orbits.  The S3 orbit, which has $r_{\rm apo}/r_{\rm peri}\sim1$ and is close to a phase space resonance, produces a stream with a velocity spread of roughly 100 km/s in the absence of substructure.  When substructure is added, the stellar debris is projected as two clumps with a slightly larger range of velocities in these coordinates.  
 Debris from the O2 orbit is, surprisingly, more compact in velocity when subhalos are present than in the smooth halo.  
 Similarly, the P3 orbit develops streams which are slightly narrower in velocity in the simulation with substructure than in the simulation without substructure.  The P1 orbit (not shown) produces stellar debris dispersed over $\sim400$ km/s in velocity at all longitudes regardless of the presence of substructure.  Thus a consistent trend towards larger velocity dispersions when substructure is present is not seen.
  
\section{Discussion}
\label{sec:disc}

The question of whether tidal streams can be used to constrain substructure in the Milky Way has been addressed by a number of studies for a variety of scenarios.  We discuss here the results of some of those works in relation to our own.

The effects of halo shape and substructure on the tidal disruption of a globular cluster are studied by \citet{ibata_lewis_irwin_quinn_02}, who find that heating due to a population of subhalos decouples from the effects of precession in a non-spherical potential by uniquely increasing dispersion in the $z$-component of angular momentum.   
Our study, while more sensitive to the effects of isolated interactions with massive substructures, is not as sensitive to this effect since, as 
noted by \citet{ibata_lewis_irwin_quinn_02}, debris from a dwarf galaxy is already fairly dispersed in $L_{z}$ and thus the effect of substructures on this quantity is more subtle.
Those authors conclude that the detection of a single cold stream from a globular cluster would provide strong evidence against the substructure picture.  In contrast, by examining a large range of scenarios we find that in some cases tidal streams from a dwarf galaxy in a halo with substructure remain as cold as streams in a smooth halo.  However, due to the differences in our approach and choice of progenitor satellite, it is unclear whether this result is generalizable to streams from a globular cluster, and thus further study is necessary to clarify the significance of detections of cold tidal streams.

\citet{johnston_spergel_haydn_02} examine the effects of substructure on test particles in a live host halo, and find that they can distinguish between smooth halos and those with substructure using their proposed `scattering index'.  Their study examines streams arising from a spherical host potential and thus does not compare the dispersion caused by substructure with that due to precession, so additional tests are necessary to determine the robustness of the scattering index for uniquely constraining substructure. Of particular importance for this study, however, is their conclusion that neglecting the response of the halo to the substructures, as we have done, is likely to underestimate the influence of the substructures due to the wake generated by their motion in the halo.  

The properties of tidal streams in an evolving host potential are examined by \citet{penarrubia_etal_06}, including a study of the effect of substructure on streams in a host potential with a spherical dark matter halo.  They conclude that substructure contributes only modestly to heating in $L_{z}$, and only see this effect in the trailing arms of streams unbound more than 5 Gyr ago, which were subjected to a higher rate of encounters with subhalos than more recently unbound material.  Our implementation of substructure does not account for the evolution of the subhalo population, and so the frequency of encounters does not have an externally set time-dependence, but depends rather on the particular orbital path.  It is demonstrated in Figure~\ref{fig:max_acc} that over $\sim 5$ Gyr a particle in our simulation will have interacted to some degree with a large number of the subhalos, and consequently we assume that our star particles will have collectively experienced all but exceptionally rare types of interactions.  Thus we expect that integrating the streams for longer times will simply enhance the features due to substructure that are already apparent.

\section{Conclusions}
\label{sec:conc}

We have investigated the effect of halo substructure on the tidal debris from a dwarf galaxy.  Our conclusions can be summarized as follows:
\begin{enumerate}

\item{Our results suggest that the shape of the host potential and the path of the orbit in question have a stronger influence on the overall spatial distribution of tidal streams than does the presence of substructure in the halo.  In light of the enormous amount of variation in the properties of the tidal debris produced by simulations in a smooth halo, it is clear that dispersal of the debris is not an effect that can be attributed exclusively, or even primarily, to substructure.  We find that the detection of a coherent tidal stream from a dwarf galaxy does not rule out the abundance of substructure considered in this work.}

\item{Substructure can noticeably shift the location of sections of debris relative to the debris in the corresponding smooth halo simulation.  This is likely to be an important complication for constructing models to fit observational data.}

\item{Interactions with a population of subhalos can lead to significant changes in the small-scale properties of the tidal debris.  Satellites on some orbits develop rich structure in the $r_{\rm sun}$-$v_{\rm los, rest}$ distribution in a smooth halo which is largely smeared out when simulated with substructure.  The presence of substructure also tends to lead to more clustered debris in a sky projection, and in some cases smaller velocity dispersions.}

\item{Stars at large distances with a small range of positive line-of-sight velocities with respect to the rest frame of the Galaxy which increase with increasing distance may be a unique signature of an interaction with a relatively massive halo substructure.  We find that particles in this configuration typically have experienced a single strong encounter with a subhalo of mass $\gtrsim 10^{10}$ M$_{\odot}$.  The formation of this feature thus depends on the orbits of the most massive subhalos in relation to the spatial and velocity distribution of stellar material.  The large number of tidal streams expected to be present in a halo like that of the Milky Way, coupled with the fact that stars from the stellar halo of the Galaxy would similarly be susceptible to such interactions, implies a high likelihood for producing these features.  Despite the fact that these particles are observed at large radial velocities and distances, they are not unbound and will thus eventually return.  Each individual feature is therefore transient.}

\item{While satellites on resonantly trapped orbits appear to produce more coherent tidal debris than non-resonant orbits in some individual cases, we do not consistently see this trend in our simulated debris.  Similarly, we do not find that streams from satellites on resonantly trapped orbits are less susceptible to the effects of substructure.}

\end{enumerate}

With the wealth of data expected from upcoming missions, we are optimistic about the prospects of using tidal streams to gain a better understanding of the mass distribution of the Milky Way.  If the halo of the Galaxy contains a population of substructure as predicted by $\Lambda$CDM, clear signatures of its presence may be detectable in the near future.

\acknowledgements
We thank A. Kravtsov for his valuable advice and guidance over the course of this study, and we acknowledge the anonymous referee whose insightful comments have substantially improved this manuscript.  We are further indebted to D. Rudd, B. Robertson, H. Morrison, S. White, C. Shapiro, V. Pavlidou, and K. Tassis for helpful discussions relating to this work.  This research was supported by the Kavli Institute for Cosmological Physics at the University of Chicago through grants NSF PHY-0114422 and NSF PHY-0551142 and an endowment from the Kavli Foundation and its founder Fred Kavli.

\bibliographystyle{apj}
\bibliography{tidal_all_v3}

\end{document}